\documentclass[10pt]{article}

\usepackage{amsmath}
\usepackage{amssymb}
\usepackage{lineno}

\usepackage{graphicx}
\usepackage{xr}

\usepackage{cite}

\usepackage{color}

\usepackage{lmodern}

\usepackage{lmodern}
\renewcommand{\familydefault}{\sfdefault}

\usepackage[labelfont=bf,labelsep=period,justification=raggedright]{caption}

\makeatletter
\renewcommand{\@biblabel}[1]{\quad#1.}
\makeatother
\date{}

\begin{document}
% Title must be 150 characters or less
\begin{flushleft}
{\Large
\textbf{Proto-cooperation: Group hunting sailfish improve hunting success by alternating attacks on grouping prey}
}
\newline
\newline
% Insert Author names, affiliations and corresponding author email.
\textbf{James E. Herbert-Read$^{1,2,\ddagger,\ast}$, 	
Pawel Romanczuk$^{3,4,5,\ddagger}$,
Stefan Krause$^{6}$,
Daniel Str\"ombom$^{1,7}$,	
Pierre Couillaud$^{8}$,
Paolo Domenici$^{9}$,	
Ralf H.J. M. Kurvers$^{3,10}$,
Stefano Marras$^{9}$,
John F. Steffensen$^{11}$,
Alexander D.M. Wilson$^{12,13}$,
Jens Krause$^{3,4}$}
\newline
\newline
$^{1}$ Department of Mathematics, Uppsala University, 75106, Uppsala, Sweden, 
$^{2}$ Department of Zoology, Stockholm University, 10691, Stockholm, Sweden, 
$^{3}$ Leibniz-Institute of Freshwater Ecology and Inland Fisheries, Müggelseedamm 310, Berlin, Germany
$^{4}$ Faculty of Life Sciences, Humboldt-Universit\"at zu Berlin, 10115 Berlin, Germany,
$^{5}$ Department of Ecology and Evolutionary Biology, Princeton University, Princeton, 08544 New Jersey, USA,
$^{6}$ Department of Electrical Engineering and Computer Science, L\"ubeck University of Applied Sciences, 23562 L\"ubeck, Germany,
$^{7}$ Department of Biology, Lafayette College, Easton, 18042, Pennsylvania, USA,
$^{8}$ D\'epartement de la Licence Sciences et Technologies, Universit\'e  Pierre et Marie Curie, 75005 Paris, France,
$^{9}$ IAMC-CNR, Istituto per l{'}Ambiente Marino Costiero, Consiglio Nazionale delle Ricerche, Localit\`a Sa Mardini, 09170 Torregrande, Oristano, Italy,
$^{10}$ Center for Adaptive Rationality, Max Planck Institute for Human Development, 14195, Berlin, Germany,
$^{11}$ Marine Biological Section, University of Copenhagen, Helsingor, 3000, Denmark,
$^{12}$ School of Life and Environmental Sciences, University of Sydney, Sydney, NSW, Australia 2006   
$^{13}$ School of Life and Environmental Sciences, Deakin University, Waurn Ponds, Victoria, 3216, Australia,
$\ddagger$ These authors contributed equally to this study,
$\ast$ E-mail: Corresponding author james.herbert.read@gmail.com
\end{flushleft}

%\linenumbers
% Please keep the abstract between 250 and 300 words
\section*{Abstract} We present evidence of a novel form of group hunting. Individual sailfish (\emph{Istiophorus platypterus}) alternate attacks with other group members on their schooling prey (\emph{Sardinella aurita}). While only 24\% of attacks result in prey capture, multiple prey are injured in 95\% of attacks, resulting in an increase of injured fish in the school with the number of attacks. How quickly prey are captured is positively correlated with the level of injury of the school, suggesting that hunters can benefit from other conspecifics' attacks on the prey. To explore this, we built a mathematical model capturing the dynamics of the hunt. We show that group hunting provides major efficiency gains (prey caught per unit time) for individuals in groups of up to 70 members. We also demonstrate that a free riding strategy,  where some individuals wait until the prey are sufficiently injured before attacking, is only beneficial if the cost of attacking is high, and only then when waiting times are short. Our findings provide evidence that cooperative benefits can be realised through the facilitative effects of individuals' hunting actions without spatial coordination of attacks. Such `proto-cooperation' may be the pre-cursor to more complex group-hunting strategies.

\section*{Introduction}
Group hunting is a fascinating example of social behaviour that can be observed in taxonomic groups including arthropods \cite{wilson1958beginnings,duncan1994group,harwood2013differences}, fishes \cite{handegard_dynamics_2012, lonnstedt2014lionfish,strubin2011group,vail2013referential,vail2014fish}, birds \cite{hector_cooperative_1986,ellis1993social} and mammals \cite{dechmann2009experimental, scheel_group_1991}. The level of coordination between individuals during hunts, both within and between these taxa, varies considerably. In its simplest form, group hunting involves hunters attacking prey together with little or no coordination of attacks while the most complex form, collaborative hunting, involves individuals adopting specific hunting roles to herd and catch their prey \cite{bailey_group_2013,gazda_division_2005,stander1992cooperative}. 

Explaining the origins and maintenance of group hunting, however, remains unresolved. Despite group hunting allowing some species to catch considerably larger prey \cite{creel_communal_1995,bailey_group_2013,anderson_teamwork_2003} as well as increasing the likelihood of making a kill \cite{creel_communal_1995}, individuals do not necessarily increase the amount of prey they consume when hunting together (compared to when hunting alone). For example, food intake per individual wolf (\textit{Canis lupus}) can be lower in larger packs compared to smaller hunting groups or lone individuals \cite{schmidt1997wolf}, and lions (\textit{Panthera leo}) do not always hunt in group sizes that optimise the amount of prey they consume \cite{packer1990lions}. Other reasons, therefore, may explain the existence of group hunting in some taxa. For example, individuals in groups may be better at limiting the access of kleptoparasites to the kill, may travel less distance, and may have a reduced likelihood of being injured during group hunts, compared to when hunting alone \cite{creel_communal_1995,vucetich2004raven,carbone1997feeding}. 

When hunters attack smaller grouping prey, the reasons for group hunting appear clearer. In some cases, group hunters use their superior speed and coordinated attacks to disrupt and fragment prey groups \cite{major_predator-prey_1978,eklov_piscivore_1994,handegard_dynamics_2012}. Groups of piscivorous fish, for example, have a higher probability of breaking up prey schools and capture more prey than single attackers \cite{major_predator-prey_1978}. Groups of humpback whales (\textit{Megaptera novaeangliae}) employ bubble-nets to capture schooling fish \cite{clapham_humpback_2000,leighton_trapped_2004} and various dolphin species have been described to use cooperative hunting strategies \cite{wursig_delphinid_1986, anderson_teamwork_2003}. Raptors have similarly been observed to use spatially-coordinated attacks to hunt flocking passerines \cite{hector_cooperative_1986, ellis1993social}. Spatially coordinated attacks appear to break down the collective defences of grouping prey, thereby increasing consumption rates for group hunters. But how did these more complex coordinated attacks evolve from simpler forms of group hunting?

In their simplest form, apparent group hunts may simply be a byproduct of clumped prey distribution, when hunters join others by eavesdropping on the cues produced from hunters finding ephemeral food patches \cite{dechmann2009experimental}. Cattle egrets (\textit{Bubulcus ibis}), for example, aggregate where prey are highly abundant, with feeding rates and prey density being closely linked \cite{scott1984feeding}. In these cases, it is unclear whether the presence of other hunters benefits individuals' hunting success. In other cases, the presence of other hunters can increase hunting success, even though hunters' attacks are not coordinated in space. Lionfish (\textit{Dendrochirus zebra}) alternate attacks on schooling prey and catch more prey when hunting in pairs than when alone \cite{lonnstedt2014lionfish}. Group hunting in a weakly electric fish (\textit{Mormyrops anguilloides}) does not appear to be spatially coordinated, and instead hunters may benefit from prey fleeing in their direction when prey escape another hunter's failed attack \cite{arnegard2005electric}. Black headed-gulls (\textit{Larus ridibundus}) capture twice as many fish when hunting in groups of six than when hunting alone, even though attacks are uncoordinated \cite{gotmark1986flock}. If individuals can benefit from the hunting actions of others without spatial coordination of attacks, then these group hunts could explain the origins of more complex group hunting strategies. But the mechanisms allowing increased capture rates for individuals with uncoordinated attacks remain unclear. One possibility is that the alternation of attacks gives hunters the opportunity to save energy, whilst others exhaust and injure the prey. This could allow individuals to benefit from increased capture success during later attacks if it is easier to catch tired, injured prey. Here we investigate whether such a 'proto-cooperative' strategy could benefit individual hunters in groups. We investigated group hunting sailfish (\textit{Istiophorus platypterus}) that alternate their attacks on schooling sardine prey (\textit{Sardinella aurita}) \cite{domenici_how_2014,marras2015not}. Attacks by sailfish appear to be uncoordinated in space, and one sailfish will abandon its attack if another individual attacks the school at the same time.

We first used behavioural observations and image analysis to systematically quantify the group hunting strategies of sailfish, which can only be done in the wild. This puts strong constraints on the type and quantity of data we could record. Therefore, to complement our empirical work, we used a mathematical model to test whether the attack-alternation strategy we observed could be effective at allowing predators to increase their capture success beyond that possible for a solitary sailfish.  We hypothesised that group hunting would allow individuals to capture more sardines per unit time using this strategy compared to if they hunted alone. Further, we evaluate the predator group sizes where this attack-alternation strategy is beneficial over solo hunting under different hunting conditions.  We also investigate whether this form of group hunting is likely to be exploited by free riders, i.e. individuals that wait until the school is sufficiently injured before attacking.

\section*{Empirical Materials and Methods}
Research was conducted 30-70 km offshore Cancun in the Gulf of Mexico (N21 28.3
-41.15 W86 38.41 -41.30). We observed group hunting sailfish separating smaller schools of sardines from larger ones containing thousands of fish. The sailfish then herded these smaller schools to the surface where the last stage of the hunt occurred. Under snorkel, we used Casio EX-FH100 cameras (operating at 240 fps) to record these smaller sardine schools that were being attacked by the sailfish. We visited this site once a year for 5 years to record the hunting behaviour of sailfish. However, we could only perform the school injury analyses (see below) in videos when sky conditions were overcast (because we required the light to be evenly distributed across the schools). This restricted the amount of data we could use. In total, our analyses are based on 63 minutes of video from 2012 documenting these interactions. Since we did not observe some of the behaviours and could not calculate some of the measures for all schools (n = 8 in total), we report the number of schools used in each analysis below. All research was conducted in line with the laws and legislation of Secretaría de Medio Ambiente y Recursos Naturales, Mexico. 

\subsection*{Attack and capture rates} During an attack, sailfish use their rostra to facilitate prey
capture by slashing or tapping the sardines \cite{domenici_how_2014}.
From the videos, we recorded the number of these attack events (n = 210 attacks across all schools) as
well as the number of successful prey capture events (n = 51 across all videos). By dividing the total number of
captures by the length of the video recordings we had recorded of particular
schools, we determined a capture rate for each school (n = 7 schools; note, we did not observe any attacks on one of the schools we recorded). We recorded the number of sardines that the bill hit during these attacks (taps or slashes). This
represents the minimum number of fish hit during these attacks because some hits may not have been visible from the
camera angle. In 52 out of the 210 attacks, we could not see how many fish
were hit and these events were excluded from analysis. We
also determined whether we could see if some of the sardines' scales were removed
during the attacks. Scale removal indicates injury to the sardines (Movie 1).
Sometimes it was not clear whether scales were removed or not during an attack (due
to subsequent obstruction by other fish), and therefore these
ambiguous events (n = 61 attacks out of 210) were not included when calculating the
proportion of fish that were injured during an attack. 

\subsection*{Proportion of the school that was injured} We investigated what proportion of each
school had injuries. We selected 39 video stills 
where light contrast across the schools was minimal. For this analysis, we only
selected multiple images from the same video if there was at least 1 minute between
the two frames of interest (see Table S1 for the number of images used for each
school). The marks on the sardines caused by injuries from the bill have a distinctive
white/pinkish appearance, different from other parts of the fish's bodies or surrounding water
(Fig. \ref{fig:exp}a). This allowed us to use image analysis to determine how injured the fish were in the school. To perform this analysis, we marked a polygon
around the edge of the school, and then cleared all pixels from outside the marked
polygon (setting their grey-scale intensity to 255). We then adjusted the brightness and
contrast of each image so that only the injury marks on the fish became pronounced.
By adjusting the brightness and contrast for each image appropriately (Fig. \ref{fig:exp}b), we
could then binary threshold the images to reveal the pixels in each image where the
injuries had occurred (Fig. \ref{fig:exp}c). Note that because the average intensity of each image
differed, we had to adjust the brightness and contrasts of each image manually. We
imported the binary converted images into MATLAB (2012b). Each image was
represented by a matrix where cells equal to zero (black pixels) were injured parts of
the school, and cells equal to 255 (white pixels) were uninjured parts of the school. By
counting the number of values in the matrix equal to zero, and then by dividing this
total by the area of the school calculated by the polygon in ImageJ, we determined the
proportion of pixels in the school depicting injuries. We determined
the mean proportion of injuries of a school if that school had been measured
in multiple images (see Table S1). We note that this semi-automated analysis provides
information on a general level of injury, which combines both the frequency and
severity of injury into one variable. 
\vspace{2ex}

\section*{Empirical Results}

\subsection*{Sailfish Group Hunting Behaviour}
Sailfish were observed in groups of $\sim$ 6 - 40 individuals hunting sardine schools (n = 8) that differed in number from approximately 25 to 100 - 150 fish (Fig. \ref{fig:SI_expdata}a; see Supplementary section 1.1). Due to observational limitations, we could not determine the exact number of sailfish that were hunting each sardine school.  Different sailfish alternated their attacks on the sardine schools (Movie S2). Individual identification of all sailfish was not possible, and therefore we could not determine the order in which individual sailfish attacked the prey school. The median time between consecutive approaches by different sailfish was 6.5 seconds (Fig. \ref{fig:SI_expdata}b; see Supplementary section 1.2). There was no relationship between the time between approaches and the sardine school size (Spearman Correlation; $\rho$ = 0.11, n = 7, P = 0.84). The median length of individuals' attacks was 2.6 seconds, but again, this was not related to school size (Spearman Correlation; $\rho$ = -0.21, n = 7, P = 0.66). A sailfish's attack was interrupted by another sailfish 19\% of the time, after which either one or both sailfish would abandon their attack.  

%%%%%%%%%%%%%%%%%%%%%%%% FIGURE %%%%%%%%%%%%%%%%%%%%%%%%%%%
\begin{figure*}
\begin{center}
\includegraphics[width=\textwidth]{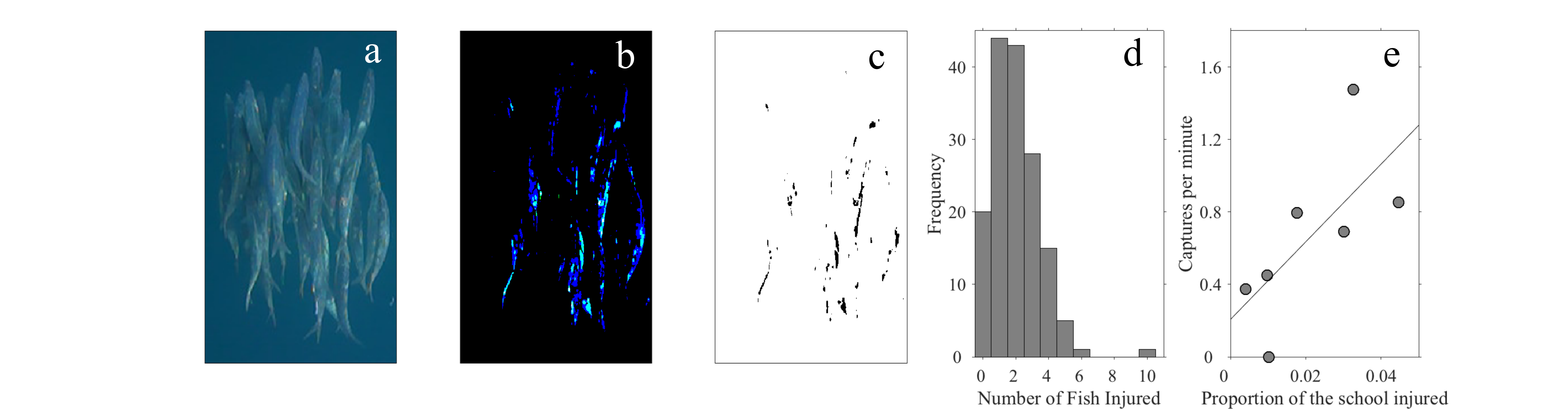}
\caption{ (a) Original frame taken from one of the videos showing the white injury marks on the sardines caused by damage from sailfish bills. (b) The original image has its contrast and brightness adjusted before binary thresholding (c), which reveals the injuries on the sardines. (d) The minimum number of sardines that were hit with the bill during a sailfish's attack (determined for 158 attacks). (e) Relationship between the proportion of the school that was injured and the sailfish capture success rate on the prey schools. Solid line represents the least squares regression; y = 21.4x + 0.21). 
	\label{fig:exp}}
\end{center}
\end{figure*}
%%%%%%%%%%%%%%%%%%%%%%%%%%%%%%%%%%%%%%%%%%%%%%%%%%%%%%%%

During attacks, sailfish used their bills to `tap' or `slash' at the sardines in an attempt to capture individual fish (Movie 1)\cite{domenici_how_2014}. Only 24\% of these attacks resulted in a successful capture and we never observed a sailfish to handle or ingest two or more fish at once. However, both the mean and the median number of sardines that were hit with a sailfish's bill during an attack was 2.0 (Fig. \ref{fig:exp}d). Whilst the attacks with the bill were very rarely observed to kill the sardines outright, sardines' scales were removed when contact was made between the sailfish's bills and the sardines' bodies in 95\% of cases. Because more fish were injured per attack than were caught, this led to many sardines in the schools having pronounced injuries on their bodies, accumulated from past attacks (Fig. \ref{fig:exp}a).  The most heavily damaged fish had over 20\% of their body covered in injuries (Fig. \ref{fig:SI_expdata}c and Supplementary section 1.3). 

We observed successful captures (n = 51) on six of the eight sardine schools we recorded. Sailfish caught individual sardines at an average rate (across all the schools) of 0.66 $\pm$ 0.17 SE sardines per minute. By quantifying the proportion of the school that was injured (Fig. \ref{fig:exp} b\&c), we found a positive correlation between the school's injury level and the capture rate (Spearman Correlation; $\rho$ = 0.82, n = 7, P = 0.03; Fig. \ref{fig:exp}e); sardines in more injured schools were captured more quickly. Given the observational constraints, we could not determine whether it was the most injured fish in the shoal that were captured next, however, we often observed injured sardines breaking off from the prey schools that presumably could not keep pace with the school. These individuals were quickly captured by the sailfish (Movie S3). There were non-significant negative correlations between capture success rates and school size (Spearman Correlation; $\rho$ = -0.54, P = 0.24, n = 7), and between the proportion of the school that was injured and school size (Spearman Correlation; $\rho$ = -0.69, P = 0.07, n = 8; see the Supplementary sections 1.4 and 1.5 for a discussion of these results).  

\section*{Group Hunting Model}
From the empirical information above, it appears that sailfish increase their capture rates as prey become progressively injured from previous attacks. But this does not explain why they hunt in groups, as a solitary hunter could get these benefits by hunting on its own. To better understand why sailfish hunt in groups, therefore, we built a simple mathematical model to capture the dynamics of the hunt. We chose to model group hunting using a non-spatial, individual-based model. On the one hand, this model effectively accounts for the fundamental temporal ``mechanics'' of the hunt observed in the field, and on the other hand is open to a full analytical investigation of its dynamics. Our model allows us to systematically investigate the rates at which sailfish catch sardines in different predator group sizes. It also allows us to explore potential differences in the strategies predators could use during the hunt. A general advantage of an individual-based approach is that our model can be easily extended to incorporate more additional features, such as agent heterogeneity or stochastic effects.  

We consider a group of $N = const.$ predators (sailfish), hunting a group of initially $S(t = 0) = S_0$ prey (sardines). The number of sailfish observed hunting in groups was $N=6-40$, however, group sizes exceeding $50$ individuals have been previously reported. Therefore in our model we studied a range of group sizes from solitary hunters $N=1$ up to a group size $N=100$. The prey schools from our empirical observations ranged from $25-150$ sardines. However, we have no information about the number of sardines that were initially separated from the school containing thousands of fish during the initial stages of the hunt. In our model calculations, therefore, we set the initial number of sardines to be larger, but in the same order of magnitude, as the largest groups observed: $S_0=200$. 

Basic biomechanics predict that small prey (sardine average body length is $\sim$  19 cm \cite{domenici_how_2014}) are more manoeuvrable than larger predators \cite{webb_locomotion_1990,domenici_scaling_2001} (sailfish are $\sim$ 200 - 250 cm). If the 
sardines can perform one or more sharp turns, removing a sailfish's potential for attack, then a sailfish is likely to abandon its attack due to its lower manoeuvrability. Meanwhile, this gives another sailfish an opportunity to initiate its attack sequence. In our model, therefore, each predator needs a finite time to perform an attack, $\tau_a$, and after an attack it requires a finite time, $\tau_r$, to be ready for the next attack. $\tau_a$ represents the time where an individual sailfish ``monopolises'' the prey school by performing its approach, manoeuvre and attack. Here, we set it to the median attack time observed for hunting sailfish; $\tau_a=2.6s$. $\tau_r$ describes the average time required by an individual hunter to prepare for the next attack sequence, i.e. for the sailfish to assume a suitable position at the rear of the prey school. This time is not available from our observations, as it requires repeated observations of a solitary sailfish hunting a sardine school. However, a reasonable time scale can be estimated from qualitative observations of the hunting process and other time scales, as well as from the assumed cooperative benefits of the hunt. Here we reasonably assume that $\tau_r$ is larger than $\tau_a$, and significantly shorter than $1$ min. Therefore, we use as a default parameter $\tau_r=20 s$. Note that whilst the attack and preparation times may vary, only their average values, $\tau_a$ and $\tau_r$, determine the conditions where group hunting is beneficial (see Fig. \ref{fig:SI_nc_vs_ta_tr} for an exploration of how $\tau_a$ and $\tau_r$ determine these conditions).

A single predator requires the time $\Delta t_{single}=\tau_a + \tau_r$ for a full attack cycle: ``perform attack'' ($\tau_a$) and ``prepare for next attack'' ($\tau_r$). Thus it attacks on average only once during this time interval and the number of attacks scales linearly with time $n_a(t)=t/\Delta  t_{single}$. If we have more than one predator, the average time interval between two attacks by a focal 
individual depends on the number of predators $N$, whereby two cases have to be distinguished: (i) If $N$ is small, then on average all other hunters can perform their attacks within the time required by the focal individual to prepare for the next attack, and the average time interval between initiation of subsequent attacks for the focal individual is simply $\tau_a+\tau_r$. (ii) If $N$ is larger than $1 + \tau_r /\tau_a$, then the focal individual will typically have to wait until other, better prepared individuals have performed their attacks. If we assume that at any time the individual with the longest waiting time will attack next, then typically all other hunters will perform their attacks between two subsequent attacks of the focal individual and the corresponding time interval becomes $N\tau_a$. In summary, therefore, the average waiting time between two attacks for an individual predator can be expressed as:
	$$\Delta t_{single}(N)=
	\begin{cases}
	{\tau_a+\tau_r} & \quad \text{for}\quad N\tau_a \leq (\tau_a+\tau_r) \\
	N\tau_a & \quad \text{for}\quad N\tau_a > (\tau_a+\tau_r) \end{cases}\ , $$
 
Using this we can calculate the average number of attacks $n_{a,s} (T,N)$ an individual predator performs until time $T$ in a group of size $N$ (Fig. \ref{fig:SI_model_na_nc}). Note that, $T$, can be interpreted simply as the time available for hunting, and is different from the actual time required to hunt down a school of sardines, $T_{tot}$ (see Fig. \ref{fig:SI_Ttot} for an exploration of how $T_{tot}$ changes depending on the hunters' group size). The average number of attacks performed by single hunter at time $T$ in a group of $N$ is given by:
		$$n_{a,s}(T,N)=\left\lfloor \frac{T}{\Delta t_{single}(N)}  \right\rfloor
		=
		\begin{cases}
			\lfloor \frac{T}{\tau_a+\tau_r} \rfloor & \quad \text{for} \quad N\tau_a \leq (\tau_a+\tau_r) \\
	        	\lfloor \frac{T}{N\tau_a}       \rfloor & \quad \text{for} \quad N\tau_a > (\tau_a+\tau_r) \ .
		\end{cases}
		$$
\noindent Here $\lfloor\cdot\rfloor$ indicates the floor function as $n_{a,s}(T,N)\in\mathbb{N}$;
\vspace{2ex}

An attacking predator has the probability $p_c$ to catch 
a single prey. During each attack, there is also a chance that prey are injured. Since it is unclear
how the injuries are distributed among individuals, we introduce a global measure of injury in the prey
school $I$. In the empirical data, we found a correlation between the level of injury of a school and the capture success rate (see Results).  This suggests that the capture probability is a monotonically increasing function of the injury level in the prey school, $p_c = p_c (I)$ (Fig. \ref{fig:model}a). The capture probability $p_c(I)$ can never exceed 1, thus it has to approach $p_{max} \leq 1$ for $I \to \infty$. Using this, and assuming that the global injury level increases linearly with the number of attacks, we may rewrite the probability of capture as a function of the number of attacks $n_a$: $ p_c(I)$ $\rightarrow$ $p_c(n_a)$ (see Fig \ref{fig:model}a, and Supplementary section 2.1). We have also checked that a nonlinear dependence of  $I$ on $n_a$ does not qualitatively affect our findings (see Supplementary section 2.2).  

During the full cycle of the focal individual (``attack sequence''+``preparation/waiting time'') on average $N$ attacks take
place, which increases the injury of the prey and therefore the capture probability. The number of attacks performed by all hunters can thus be expressed as $n_a=i N$, with $i$ being the number of the attacks by a focal individual and $N$ being the group size. We can calculate the expected number of prey captured by a focal 
predator at time $T$ by summing over all the capture probabilities $p_c(i  N)$ during its subsequent attacks $i$.
Here we have to take into account that the total school size imposes an upper limit on
the possible number of fish caught, which is simply the average number of prey per
predator $S_0/N$. Thus the expected number of prey captured per predator is:

%%%%%%%%%%%%%%%%%%%%% EQUATION %%%%%%%%%%%%%%%%%%%%%%%%%

\begin{align}
	\label{eq:caught_vs_N}
	\langle n_c \rangle(T,N) = \text{min}\Big\{\sum_{i=0}^{n_{a,s}(T,N)} p_c(i N),\  \frac{S_0}{N} \Big\}
\end{align}

with $n_{a,s}(T,N)$ being the number of attacks performed by the focal individual in a group of $N$ hunters, up to $T$ (the time available for hunting). We have explored the model's behaviour with different parameters, and whereas quantitative results might differ, the overall results appear surprisingly robust and the qualitative predictions remain unchanged.

\subsection*{Group Hunt Simulations}
In order to test our theoretical predictions, we performed numerical simulations of a simple individual-based model. $N$ hunters perform subsequent attacks on a school of $S(t)$ sardines, with the initial school size being $S(t=0)=S_0$. The attack sequence of each hunter has a fixed duration $\tau_a$. The attack may lead to a successful capture of a single prey with probability $p_c(n_a)$, which is a function of the number of all previous attacks on the school according to Eq. S2 in the Supplementary. A simulation run is terminated when all prey are captured $S(t)=0$. The preparation time for the next attack for each hunter is $\tau_r$. The initial attack order is set randomly. As time progresses, the next attack is performed by the individual with the longest waiting time. For a fixed attack duration $\tau_a$ and preparation time $\tau_r$, the initial attack order of the hunters remains unchanged. All results are obtained by averaging $100$ independent simulation runs. We confirmed that our results are robust with respect to random attacks and preparation times, which introduces additional stochasticity and randomises the order of the hunters within a single run (see Supplementary section 2.3 for details).   

\subsection*{Are there benefits for free riders?}
This form of group hunting immediately raises questions surrounding the existence of producers and free riders in groups. Producers (hunters that begin attacking from the start of the hunt) generate a public good where higher levels of prey injury leads to higher capture success rates. There is the potential, therefore, for free riders (individuals that delay their attacks for some time until the school is sufficiently injured) to avoid paying the costs of attacking at the beginning of the hunt where the initial capture probability is low, and profit from the higher capture probability at later stages of the hunt. 

In order to explore possible fitness trade-offs in terms of the energy expenditure versus energy uptake, we combined the stochastic individual based model with an energetic balance equation (See Supplementary section 2.4). We consider an ``optimal'' situation of being a single free rider hunting with $N-1$ producers. The free rider refrains from attacking prey at the beginning of the hunt for a time $T_{fr}$ (attack delay time). Using the energy payoffs an individual receives during the hunt, $\Delta e_{total,i}$, we can calculate the relative energy payoff of an individual $f_i$ within a population:

\begin{align}
	f_i = \frac{\Delta e_{total,i}-\text{min}(\Delta e_{total,i})}{\text{max}(\Delta e_{total,i})-\text{min}(\Delta e_{total,i})}
\end{align}

\noindent which scales between $0$ for minimal energy payoff and $1$ for maximal energy payoff. Here, $\Delta e_{total,i}$ is a function of a sailfish's base energy expenditure (the energy needed to simply remain with the prey school), the energy required to perform attacks, and the energy received by the captures it makes during a hunt (See Supplementary section 2.4).

In order to assess possible energy benefits of free riders, we calculated the difference between the average relative energy payoff of free riders and producers 
\begin{align}
	\Delta f = \langle f \rangle_{fr} - \langle f \rangle_{prod}. 
\end{align}

Positive values indicate an advantage for the free riding strategy, whereas negative values indicate on average higher energy pay-offs for the producers. All results discussed were obtained by simulating $100$ independent runs for each group size $N$ and attack delay time $T_{fr}$. 

\section*{Modelling Results}

%%%%%%%%%%%%%%%%%%%%% FIGURE %%%%%%%%%%%%%%%%%%%%%%%%%%
\begin{figure*}
\begin{center}
\includegraphics[width=0.325\textwidth]{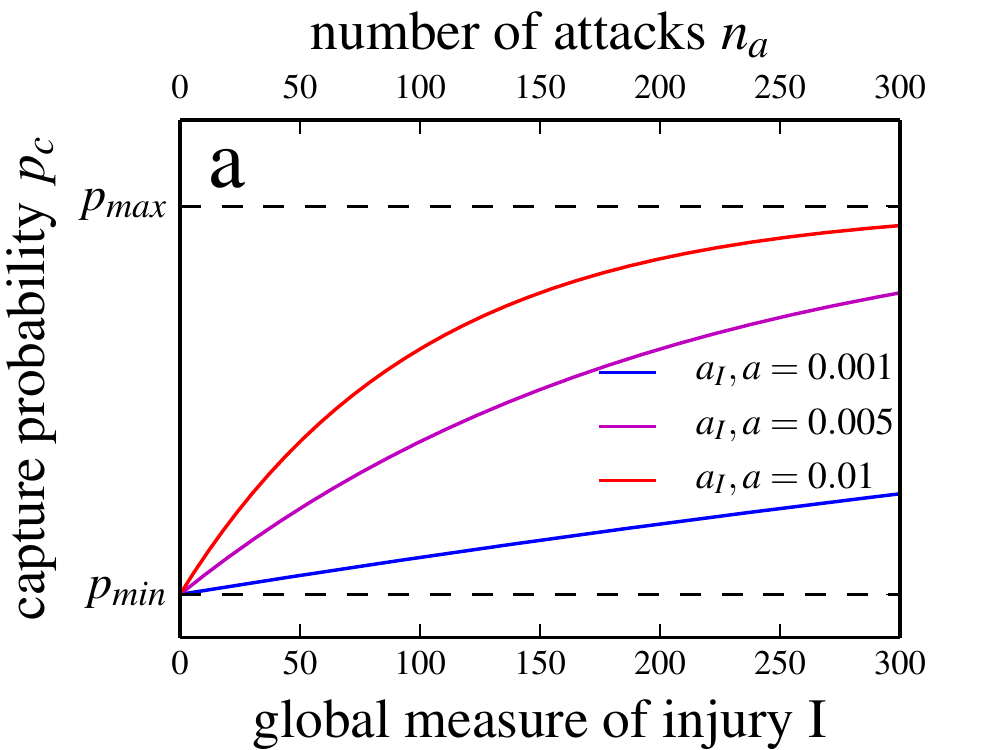}
\includegraphics[width=0.325\textwidth]{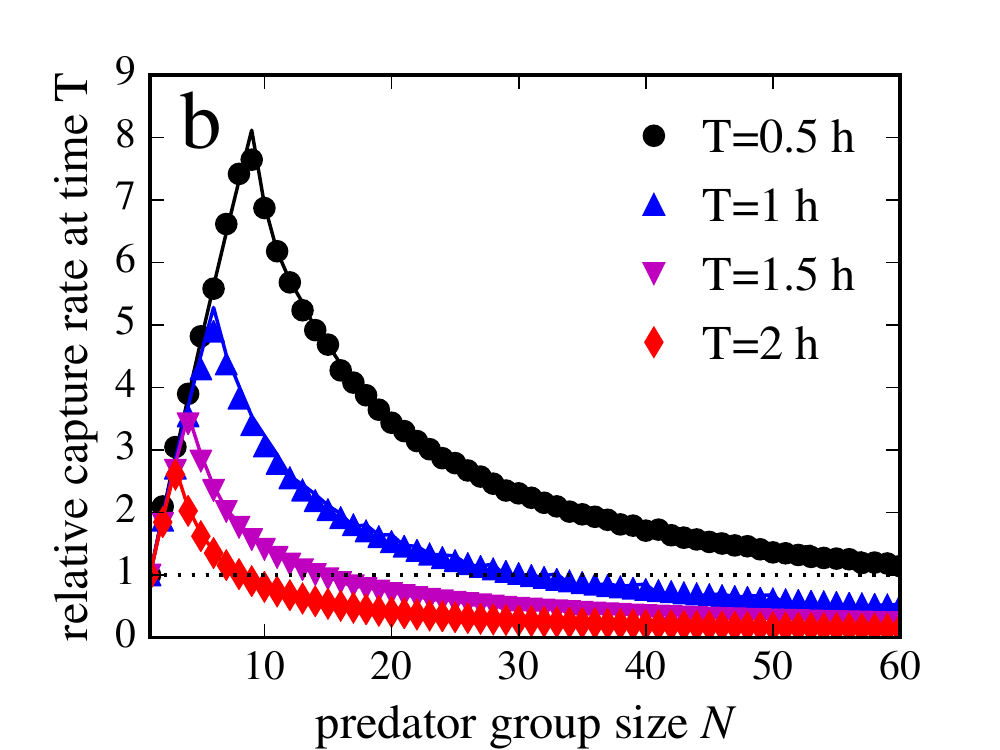}
\includegraphics[width=0.325\textwidth]{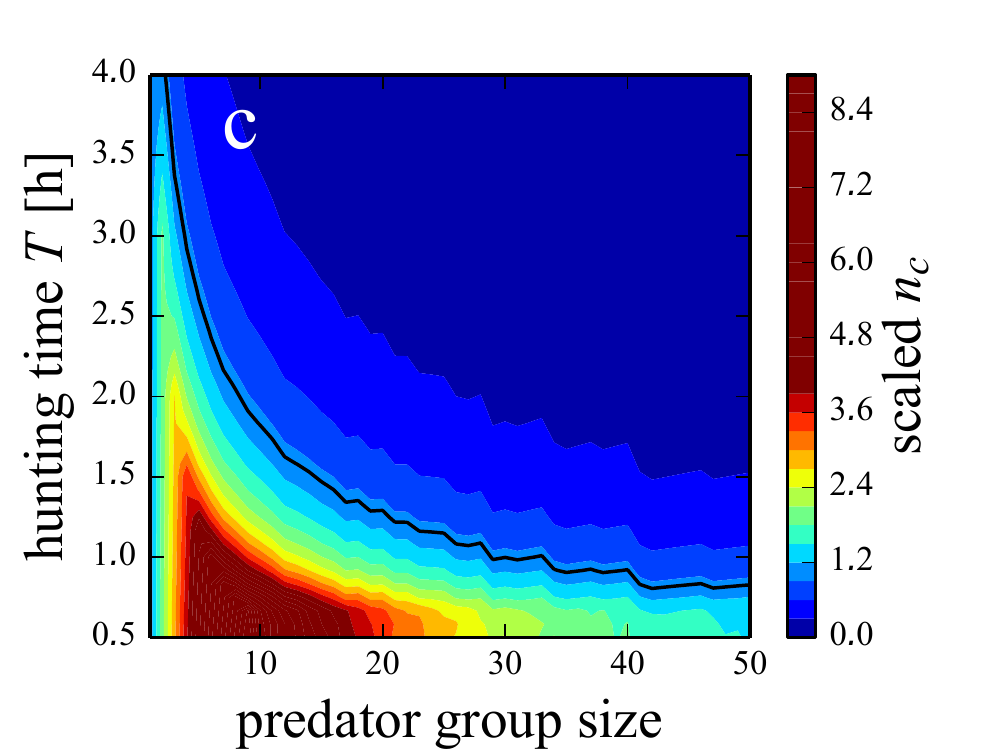}
\caption{(a) Capture probability versus the global measure of injury of the prey school and the number of attacks for different values of the injury growth constants $a_I$, $a$ (see Supplementary Eqs. \ref{eq:pc_vs_I} \& \ref{eq:pc_vs_na} for details) assuming $\Delta I=1$. The dashed lines indicate $p_{min}$ and $p_{max}$. (b) 
Group hunting model: Number of prey captured per individual as a function of $N$ scaled by the number of prey captured for a solitary
predator (horizontal dotted line). The largest group sizes which offer an advantage
to solitary hunting are typically observed for short times $T \leq 1h$ and
decrease for longer times (or small prey schools). (c) Theoretical prediction on number of fish captured per predator versus predator
group size for varying hunting times $T$. Again, $n_c$ is normalised by the number of prey
captured by a solitary predator. Solid line shows the contour corresponding to the
value of $n_c = 1$ (same as solitary hunter) and represents therefore the border of the
region where group hunting is beneficial. Default model parameters in all panels if not varied or otherwise stated: $\tau_a = 2.6s$, $\tau_r =20s$,  $a =5\cdot10^{-4}$, $p_{min} =0$, $p_{max}=1$, $S_0 = 200$.
\label{fig:model}\label{fig2}}
\end{center}
\end{figure*}
%%%%%%%%%%%%%%%%%%%%%%%%%%%%%%%%%%%%%%%%%%%%%%%%%%%%%%%%

If the time available for hunting $T\to\infty$, the expected number of sardines caught $\langle n_c \rangle$ is always equal to $S_0/N$, and always has a maximum at $N=1$. Hence if time is not a limiting factor, then it is always better for a predator to hunt alone because it would not have to share prey with conspecifics. However, predators may attempt to maximise how many prey they catch per unit time (i.e. the capture rate), and not just the absolute amount of prey they catch (see Supplementary section 1.6). Under this scenario, it may not be beneficial to hunt alone. By performing numerical simulations of the model we determined the conditions where group hunting can improve capture rates for individual sailfish. Figures \ref{fig:model}b and c show the number of fish captured per predator as a function of group size $N$, scaled by the number of prey a solitary hunter ($N = 1$) would have caught at that time (see Fig. \ref{fig:SI_nc_unscaled} for unscaled values). In this way we can identify the maximum group size $N_m$, where each individual outperforms a solitary hunter. This depends strongly on the available hunting time. Whilst the optimal group sizes that maximise prey intake rates per hunter are small (10 when hunting times are short ($T=0.5h$) to 3 when hunting times are long ($T=2h$)), the group sizes where group hunting outweighs hunting alone are typically much larger.  For $T =0.5h$ we observe $N_m =70$, which then quickly decreases to $N_m = 30$ for $T = 1h$ and $N_m=13$ for $T=1.5h$. Eventually, for $T\to\infty$, $N_m$ will always converge to 1 due to the finite size of the school. At short times $T$, $N_m$ is always larger than $1+\tau_r/\tau_a$, which is the group size at which the individual hunters start to pay temporal costs of group hunting (Fig. \ref{fig:SI_model_na_nc}). We checked how changing the initial number of sardines in the prey school, $S_0$, affected the conditions under which group hunting was beneficial. Smaller (or larger) initial prey group sizes shifted the hunting times so that shorter (or longer) times $T$ made group hunters outperform solitary hunters.

%%%%%%%%%%%%%%%%%%%%% FIGURE %%%%%%%%%%%%%%%%%%%%%%%%%%
\begin{figure*}
\begin{center}
\includegraphics[width=0.345\textwidth]{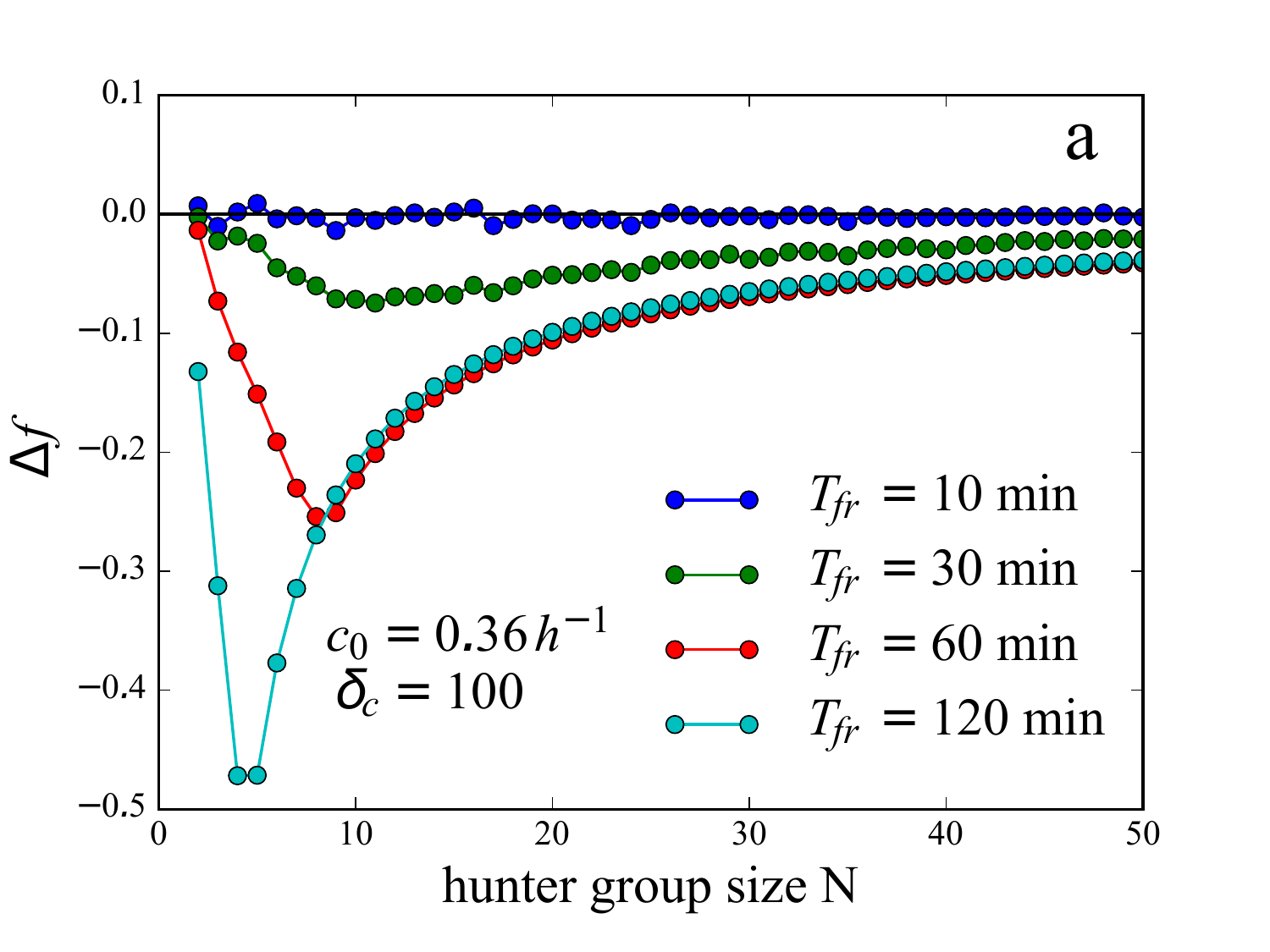}
\includegraphics[width=0.345\textwidth]{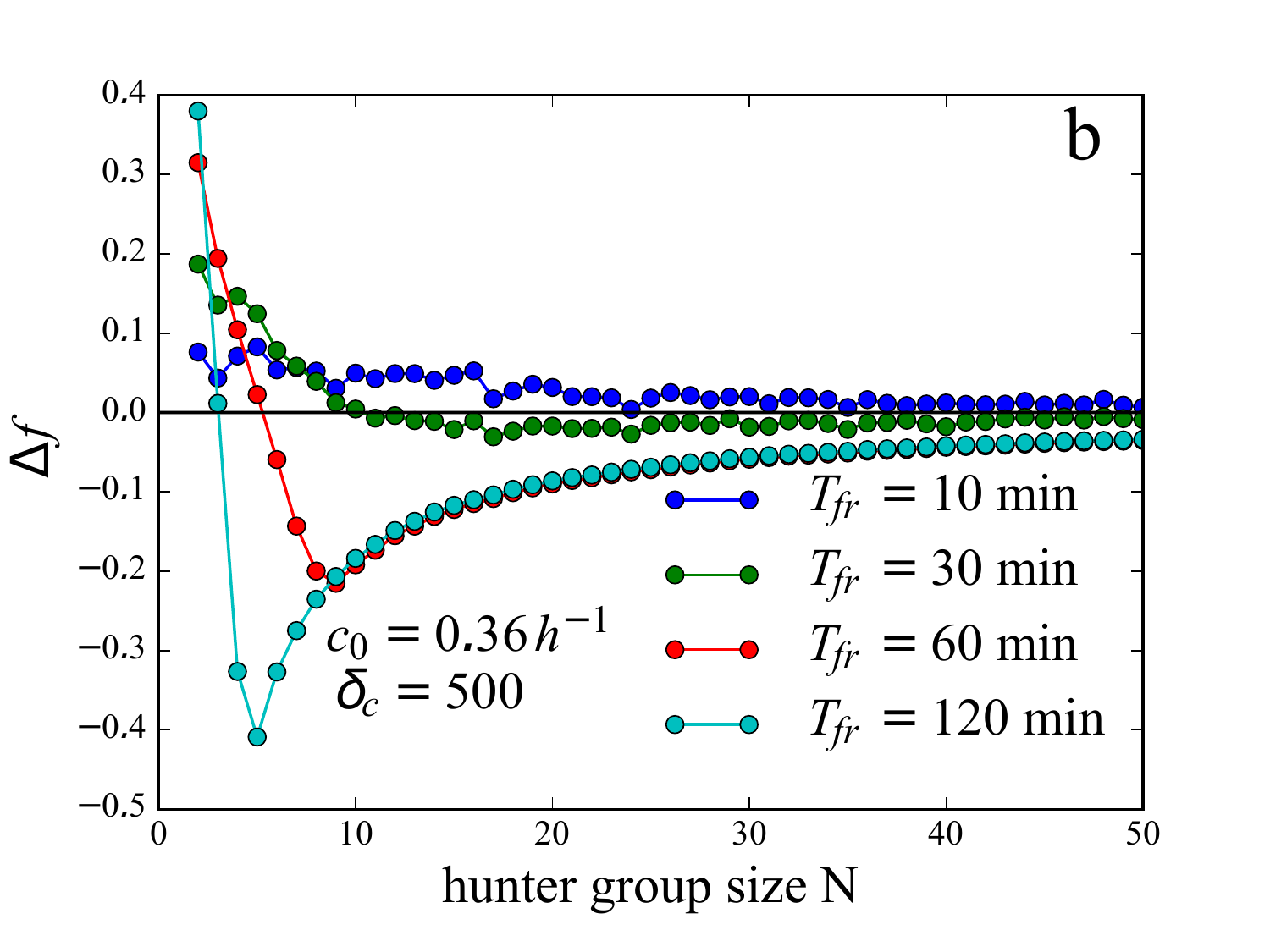}
\includegraphics[width=0.288\textwidth]{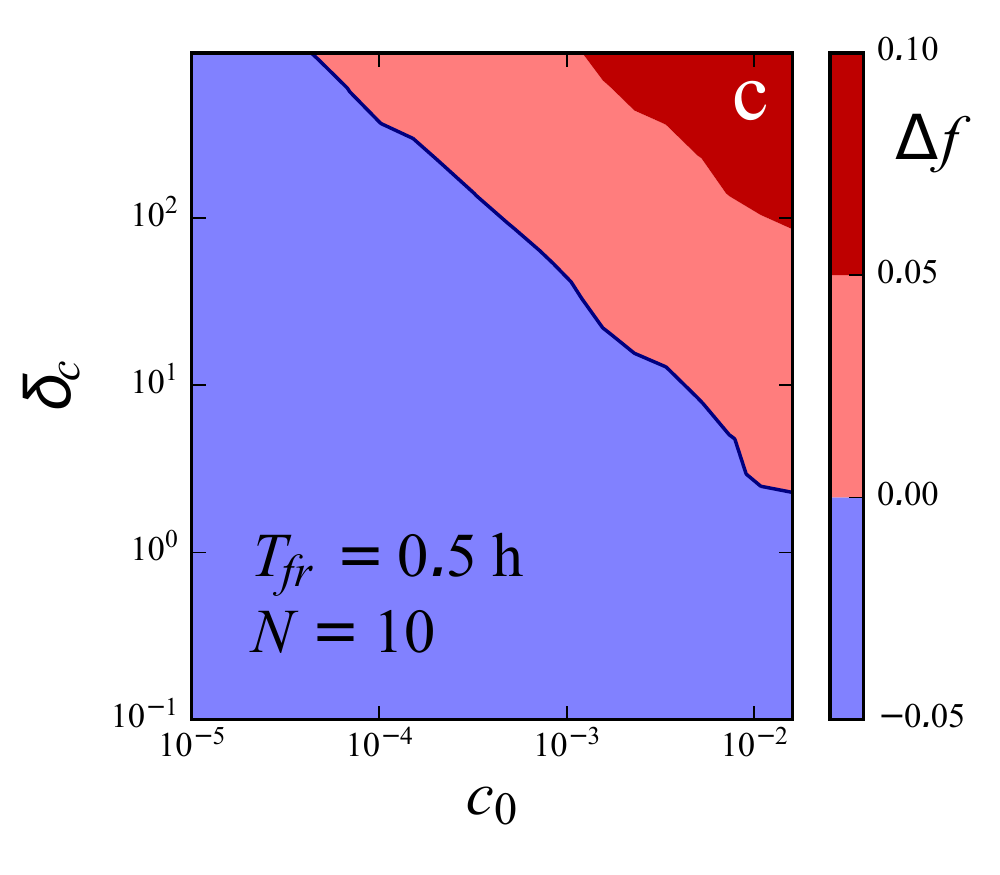}
 \caption{(a,b) Relative energy payoff difference $\Delta f$ versus hunter group size $N$ for different attack delay times $T_{fr}$ with $c_0=0.0001 \text{s}^{-1}=0.36 \text{h}^{-1}$ and different values of the relative energetic costs of attacks $\delta_c=100$ (a) and $\delta_c=500$ (b). (c) $\Delta f$ versus $c_0$ and $\delta_c$ for fixed $T_{fr}=0.5$h and $N=10$ (blue region indicates $\Delta f<0$; i.e. where free riding is not beneficial). All other simulation parameters as in the main text and Supplementary section 2.4.  
		 \label{fig:scrounge}\label{fig_3}}
\end{center}
\end{figure*}
%%%%%%%%%%%%%%%%%%%%%%%%%%%%%%%%%%%%%%%%%%%%%%%%%%%%%%%%

We also investigated whether a free riding strategy could be beneficial for some individuals in groups. One key parameter, $\delta_c$, predominantly controls whether there is an advantage for the free riding strategy with respect to energy payoffs (see Supplementary section 2.4 for details). $\delta_c$ is a dimensionless number that represents the effective increase of the energy expenditure during an attack relative to the base energy expenditure. A value of $\delta_c=1$, would correspond to a doubling of the energy consumption rate during an attack sequence.  An advantage for the free riding strategy can only be observed for very large values of $\delta_c\gg 10$ (Fig. \ref{fig:scrounge}a\&b). Even then, this advantage only becomes significant for small hunter group sizes $N<10$ or small attack delay times $T_{fr}$ (Fig. \ref{fig:scrounge}b). Decreasing values of $\delta_c$ and the base energy consumption rate, $c_0$, make free riding increasingly unlikely for a given group size and hunting time (Fig. \ref{fig:scrounge}c). Free riding remains disadvantageous for large regions of parameter space if we allow for nonlinear dependence of the injury level on the number of attacks (Supplementary section 2.2), or if their is the potential for the hunt to be interrupted (see Supplementary section 2.5). 

%%%%%%%%%%%%%%%%%%%%%%%%%%%%%%%%%%%%%%%%%%%%%%%%%%%%%%
\section*{Discussion}
We have proposed a simple mechanism that can explain why sailfish hunt in groups. During a sailfish's attack, more fish are injured than are caught. Injuries can simply result as a byproduct of the sailfish attempting to catch sardines, and we do not suggest the sailfish are attempting to injure but not catch the sardines during attacks. Sailfish bills are covered in small denticles or micro-teeth  \cite{fierstine_use_1996,domenici_how_2014}, which likely facilitates this injury. Because more fish are injured than caught per attack, this necessarily leads to more injured fish in the school as the number of attacks increases. We found a positive correlation between the injury level of a school and the rate at which prey in that school were caught. Our modelling approach demonstrated that individuals using an attack alternation strategy whilst hunting in groups can achieve increased per capita capture rates compared to if hunting alone. This strategy does not require spatial coordination of attacks between hunters. Simply hunting in a group can improve capture success rates, even though individuals do not need to change how they attack prey whether alone, or in groups.

Like other systems \cite{arnegard2005electric,gotmark1986flock}, sailfish do not appear to spatially coordinate their attacks.  In fact, sailfish predominantly attack the prey schools when no other individual is doing so (presumably to reduce the risk of individuals injuring themselves during attacks). This suggests that sailfish time their attacks to generally take place after another hunter has departed the prey school. Indeed, temporally coordinated attacks have been observed in other species that hunt grouping prey and have been shown to improve capture success rates \cite{lonnstedt2014lionfish,thiebault2016capture}. Modelling studies have also indicated that temporally coordinated attacks act to improve capture success rates for group hunters \cite{lett2014effects}. Whilst temporal, and not spatial coordination occurs during individuals' attacks, spatial coordination may occur in other aspects of the hunt. Sailfish herd and chase their prey, which may involve individuals moving to positions around the prey school that are dependent on the positions of other hunters. Alternatively, this herding behaviour may simply be a byproduct of predators occupying empty space around the prey school, without direct coordination between predators' movements, unlike other group hunting species \cite{wursig_delphinid_1986}. Fine resolution sonar data will be needed to investigate these herding dynamics further. In any case, we have demonstrated that group hunting can benefit individual sailfish without spatially coordinated attacks or individuals adopting specific hunting roles. Our results also highlight that temporal, rather than spatial coordination of attacks, may allow for simpler forms of cooperative behaviour to evolve.

We found a correlation between the level of injury of the prey school and the capture success rates of sailfish. Whilst we interpret this as a causal link, there are other explanations which could lead to this correlation. For example, if some groups of predators are more efficient at catching prey (and as a byproduct injure more sardines in the school) than other predator groups, this may lead to higher capture rates on more injured schools. Little is known about the social organisation of group-hunting marine fish \cite{freon2000review}. The traditional assumption has been that these predators live in fission-fusion systems with little social cohesion \cite{krause_social_2000}. However, novel tracking technology and interest in social networks have provided a fresh methodological and conceptual approach to this topic, producing some evidence for significant co-occurrence of particular individuals \cite{wilson_social_2014,bayliff_integrity_1988,klimley_school_1999}. Understanding the social organisation of sailfish groups, perhaps by identifying individuals using unique markings or sail patterns would greatly improve our understanding of this system.   

More work is needed to determine the causal mechanism between increased capture success and prey injury level. Whilst we observed injured fish breaking off from the shoal that were quickly captured, it may not always be the most injured fish in groups that are captured next. Injured fish may have reduced ability to transfer directional information about a predators' location, which in turn could affect the school's escape manoeuvres \cite{gerlotto2006waves,nottestad1999herring}. This may lead to non-injured fish being at greater risk in injured, versus non-injured prey schools. It is also likely that multiple attacks can have internal physiological effects on prey behaviour. For example, attacks over time are likely to reduce the energy stores in prey, reducing their ability to perform escape manoeuvres or sustain high escape speeds through fatigue. Hence sustained attacks, even without predators actively injuring their prey, could lead to increased capture success rates for group hunters \cite{domenici2000spacing,guinet2007killer}. This may explain why in other systems, prey intake rate increases as a function of group size, without predators coordinating their attacks \cite{gotmark1986flock}. Indeed, the attack and success rates of other marine predators that attack schooling prey are in the same order of magnitude as our study
 \cite{parrish1992levels,parrish1989predation}. In theory, our model can be applied to any system where the likelihood of capturing prey increases as a function of the number of attacks of previous predators.  

Cooperation through turn-taking strategies has been described in other systems, for example, in predator-inspection behaviour in fish \cite{dugatkin_cooperation_1997,milinski1987tit}. But the exploration of turn-taking has usually been assessed in dyads and the role of turn-taking is not well understood in larger groups. Indeed, turn-taking strategies in larger groups raise interesting questions regarding the potential for cheating \cite{dugatkin1997evolution}. It has previously been proposed that when hunters attack small grouping prey that cannot be shared, there is no temptation to cheat, as not participating in the hunt returns no pay-offs \cite{packer_evolution_1988}. However, this approach did not consider that it may be easier to catch prey over time as they receive injuries or become exhausted. The increase in injured prey over time can be interpreted as a public good \cite{levin2014public}, which may be susceptible to exploitation by cheaters that delay the onset of their attacks. In microbial communities with diffusing public goods, the partial monopolization of resources due to spatial localization, may promote cooperation by denying non-producers access to resources \cite{allen2013spatial,drescher2014solutions}. While such an explicit spatial effect is likely to be absent in the highly dynamic turn-taking hunting process of sailfish, a direct analogy can be drawn via the intrinsic coupling of the production of the public good and the capture of individual prey. Producers (attackers) have access to the prey school from the onset of the hunt, albeit with low initial capture probabilities. Our investigation of the potential energetic trade-offs during the hunt suggests that individuals that delay their onset of attacks (free riders) would only benefit from such a strategy if the cost of attacking was 10 times higher than simply remaining with the prey school. Future work, with observations on the behaviour of individually identifiable predators, will be needed to determine if this strategy exists. Nevertheless, opportunistic access to the prey school, combined with the byproduct of injuring prey during attempted captures, can promote individual hunting success in groups. We regard this form of group hunting, which does not require explicit cooperation, as `proto-cooperation'.

Our results demonstrate that individuals can benefit from group-hunting without specific hunting roles (as in collaborative hunting), higher social-organisation or complex cognition. Whilst hunting in groups potentially reduces the total amount of prey an individual predator is likely to catch, sailfish can offset this by collectively catching more prey per unit time when hunting together. This facilitative hunting method raises new questions surrounding the evolution of cooperative behaviour in group living animals.

\section*{Competing Interests}
We have no competing interests.

\section*{Author Contributions}
All authors except PR, DS and SK collected field data.  JH-R and DS analysed the empirical data and performed statistical analyses.  PR designed the modelling component of the paper.  All authors wrote the paper. All authors gave final approval for publication.

\section*{Data Availability}
All data accompanying this paper are available at DataDryad.com doi:10.5061/dryad.t9m6c

\section*{Acknowledgments}
We thank Rodrigo Friscione Wyssmann and the staff of Solo Buceo for their help in the field. We thank two anonymous referees, B. Taborsky, I.D. Couzin, D.J.T Sumpter, C.E. Tarnita and M. Wolf for helpful comments or discussions on the manuscript. 

\section*{Funding}
JK and PR acknowledge funding from Leibniz-Institute of Freshwater Ecology and Inland Fisheries. PR acknowledges the funding via the P.R.I.M.E. Fellowship by the German Academic Exchange Service. JH-R was supported by a Knut and Alice Wallenberg Foundation Grant awarded to D.J.T Sumpter.

%\begin{thebibliography}
%%\end{thebibliography}

\clearpage

\renewcommand{\familydefault}{\sfdefault}
\renewcommand\thesection{\arabic{section}}
\newcommand{\mean}[1]{\left < #1 \right >}
\newcommand{\abs}[1]{\left | #1 \right |}
\renewcommand{\vec}[1]{\mathbf{#1}}
\newcommand{\fourier}[2]{{\hat{#1}}_{#2}}
\newcommand{\nabcp}{\underset{\sim}{\nabla}}
\newcommand{\red}[1]{\textcolor{BrickRed}{#1}}
\newcommand{\green}[1]{\textcolor{ForestGreen}{#1}}
\newcommand{\blue}[1]{\textcolor{NavyBlue}{#1}}

\renewcommand{\thepage}{\arabic{page}} 
\renewcommand{\theequation}{S\arabic{equation}} 
\renewcommand{\thesection}{\arabic{section}}  
\renewcommand{\thetable}{S\arabic{table}}  
\renewcommand{\thefigure}{S\arabic{figure}}

\newcommand{\eqrefit}[1]{\textit{(\ref{#1})}}
\setcounter{figure}{0}

% Title must be 150 characters or less
\begin{flushleft}
{\Large
\textbf{Supplementary Material: Proto-cooperation: Group hunting sailfish improve hunting success by alternating attacks on grouping prey}
}
Proceedings of the Royal Society B: Biological Sciences, DOI: 10.1098/rspb.2016.1671
\newline
% Insert Author names, affiliations and corresponding author email.
\textbf{James E. Herbert-Read$^{1,2,\ddagger,\ast}$, 	
Pawel Romanczuk$^{3,4,5,\ddagger}$,
Stefan Krause$^{6}$,
Daniel Str\"ombom$^{1,7}$,	
Pierre Couillaud$^{8}$,
Paolo Domenici$^{9}$,	
Ralf H.J. M. Kurvers$^{3,10}$,
Stefano Marras$^{9}$,
John F. Steffensen$^{11}$,
Alexander D.M. Wilson$^{12,13}$,
Jens Krause$^{3,4}$}
\newline
\newline
$^{1}$ Department of Mathematics, Uppsala University, 75106, Uppsala, Sweden, 
$^{2}$ Department of Zoology, Stockholm University, 10691, Stockholm, Sweden, 
$^{3}$ Leibniz-Institute of Freshwater Ecology and Inland Fisheries, Müggelseedamm 310, Berlin, Germany
$^{4}$ Faculty of Life Sciences, Humboldt-Universit\"at zu Berlin, 10115 Berlin, Germany,
$^{5}$ Department of Ecology and Evolutionary Biology, Princeton University, Princeton, 08544 New Jersey, USA,
$^{6}$ Department of Electrical Engineering and Computer Science, L\"ubeck University of Applied Sciences, 23562 L\"ubeck, Germany,
$^{7}$ Department of Biology, Lafayette College, Easton, 18042, Pennsylvania, USA,
$^{8}$ D\'epartement de la Licence Sciences et Technologies, Universit\'e  Pierre et Marie Curie, 75005 Paris, France,
$^{9}$ IAMC-CNR, Istituto per l{'}Ambiente Marino Costiero, Consiglio Nazionale delle Ricerche, Localit\`a Sa Mardini, 09170 Torregrande, Oristano, Italy,
$^{10}$ Center for Adaptive Rationality, Max Planck Institute for Human Development, 14195, Berlin, Germany,
$^{11}$ Marine Biological Section, University of Copenhagen, Helsingor, 3000, Denmark,
$^{12}$ School of Life and Environmental Sciences, University of Sydney, Sydney, NSW, Australia 2006   
$^{13}$ School of Life and Environmental Sciences, Deakin University, Waurn Ponds, Victoria, 3216, Australia,$\ddagger$ These authors contributed equally to this study,
$\ast$ E-mail: Corresponding author james.herbert.read@gmail.com
\end{flushleft}

%\linenumbers

\section{Empirical Observations}

\subsection{\normalsize School size analysis} From the videos, we selected single frames of the sardine schools (n =
123 frames; 8 different schools; see Table S1 for number of images per school) to be
analysed and exported them using VirtualDub (v 1.9.11). We imported the images into
ImageJ (v 1.36b) and measured the lengths of five haphazardly selected fish in each
image (in pixels) using ImageJ's internal measure function. We marked a polygon
around the edge of the school's members and calculated the internal area of this
polygon, again using the measure function. Dividing the area of this polygon (in
pixels) by the average length of the five selected fish gave us a proxy for the relative size
of each school (Fig. \ref{fig:SI_expdata}a). Sardine length is generally uniform across
these schools  \cite{domenici_how_2014}.

\subsection{\normalsize Approach frequency}
We recorded the time between consecutive approaches by
different sailfish towards the sardine schools (n = 7 schools) (Fig. \ref{fig:SI_expdata}b). The time of approach
was determined as the time when a sailfish was within one sailfish body length of the
sardine school with its dorsal fin raised. This behaviour is typically observed before an
imminent attack \cite{domenici_how_2014,marras2015not}. If a sailfish was already approaching the sardine school at the very start of the video, we recorded the time of approach as zero. We also recorded the time at which this sailfish departed
the school, which was defined as when the sailfish swam away from the school. The time between the approach and the departure was measured as
the `attack length'. We also determined the time between one sailfish departing and
another sailfish approaching. On 19\% of occasions, one sailfish approached the school before
another sailfish had departed it. In these cases, one or both sailfish always abandoned their attack.   

\subsection{\normalsize Injuries on individual fish} We investigated the extent to which individual sardines
were injured. We sampled images from the videos where unobstructed individual injured fish could be seen. In each image, we selected 1- 2 individuals that were visibly
the most injured (n = 45). We only selected a fish if another fish obstructed less than $\sim$
10\% of its body surface. Like in the school injury analysis, but now based only on single
individuals, we drew a polygon around the outline of the focal fish, calculated the area
of the polygon and cleared all pixels from outside this polygon (making their intensity
= 255). We then adjusted the brightness and contrast of each image before binary
thresholding, and then imported these binary images into MATLAB. We then summed
the number of pixels indicating injures (values equal to zero) and divided this total by
the area of the polygon measured in ImageJ to determine the proportion of the body of
a sardine that was injured.
\vspace{2ex}

\subsection{\normalsize Group size effects}
Whilst we concentrated on how the sailfish progressively injured prey over time, and how capture rates were correlated with the level of injury in the prey school, there was also a weaker correlation between capture rates and school size (Spearman Correlation; $\rho$  = -0.54, P = 0.24, n = 7). If indeed capture rates do increase as school size decreases (and we could not detect this effect due to limited sample size), then our model could be more broadly applied to other systems. The confusion effect decreases as group size decreases, sometimes making it easier to catch prey in smaller group sizes \cite{landeau1986oddity,ioannou2008confusion}.  If individual hunters progressively decrease the size of the prey group over time, then this could allow hunters in the future to increase capture rates in subsequent attacks. Hence this model may not only apply when prey are injured or fatigued, but also when prey group size decreases over time. 

\subsection{\normalsize Injury in the school}
There was a negative trend between injury level and school size (Spearman Correlation; $\rho$  = -0.69, P = 0.07, n = 8). Whilst again, the non-significant trend could be due to limited sample size, it may also be due to the dynamics of the hunt. Sailfish break off smaller schools from larger schools numbering into the hundreds of thousands of fish.  If a small school was isolated relatively recently, it could in theory have a very low injury level compared to a larger group that had been under attack for a long time. Presumably our observed schools had variable initial sizes and attack durations that could introduce confounding variation into the levels of injury. We also note that our measure of the proportion of the school that was injured combines both the severity of injuries on individual fish, and the spread of injuries across different fish. Both the severity of injuries and spread of injuries across individuals are likely to be important in this system. We sometimes observed very injured sardines breaking off from the school, and these individuals were quickly consumed by the sailfish. Hence the level of injuries on singular fish are likely to be important for improving capture success rates as well as the general injury level of the school.

\subsection{\normalsize Why capture rates are likely to be important for group hunting sailfish}
We identified that increased capture rates per unit time was a key benefit for individual hunters in groups.  But why might these rates be important to increase, and why might hunting time be restricted? On two occasions we observed spotted dolphins, (\textit{Stenella attenuata}), arriving at the sardine school that were under attack by sailfish. On arrival at the sardine school, the dolphins used their tails to stun and disperse the whole school. Sailfish that have potentially invested hours into injuring and exhausting their prey can thus lose their fish to the kleptoparasitic dolphins in a few seconds. The number of daylight hours is also likely to put an upper limit on available hunting time. Studies on tagged sailfish show that their time spent near the surface increases during the day (compared to the night) \cite{hoolihan2005horizontal}, which suggests that hunting primarily takes place at the surface during daytime periods. Hunting time is also likely to be constrained because once the targeted school has been consumed, the sailfish have to find another larger school, separate off a smaller school, and begin hunting again.  Reducing the time between these events is presumably important in a time limited system. Further, because individual prey items are not shared, prey caught per unit time may be a particularly important measure of success during these hunts. 

\section{Modelling Group Hunting}

\subsection{\normalsize Mathematical model - capture probability $p_c$}
\label{sec:pc}
We assume that the capture probability $p_c$ is a monotonically increasing function of the global injury level $I$, and is bounded by $p_{min}$, $p_{max}$: $0\leq p_{min}\leq p_c(I)\leq p_{max}\leq1$. There are infinitely many functions that fulfil these requirements, and based on our empirical observations, we have no \emph{a priori} arguments to choose a particular one. However, the qualitative results will be independent on the particular choice of $p_c(I)$.   
%%%%%%%%%%%%%% CUT FROM MAIN%%%%%%%%%%%%%%%%%%%%%%%%%%%%%%%%%%%%%%%%%%% 
Here we choose

%%%%%%%%%%%%%%%%%%%%% EQUATION %%%%%%%%%%%%%%%%%%%%%%%%%
\begin{align}\label{eq:pc_vs_I}
	p_c(I)=p_{min}+(p_{max}-p_{min})(1-e^{-a_I I})
\end{align}
%%%%%%%%%%%%%%%%%%%%%%%%%%%%%%%%%%%%%%%%%%%%%%%%%%%%%%%
which increases linearly with $I$ for small injury levels and approaches $p_{max}$ asymptotically for $I\to\infty$ (see Fig. \ref{fig:model}a, main text). This function naturally fulfils the monotonic increase in injury level with a necessary saturation level. We have also checked whether a nonlinear dependence of $n_a$ on $I$ affects our qualitative findings (see below, Sec \ref{sec:nonlinear}). It is impossible to obtain a reliable value for the initial probability to catch a prey during a full attack sequence (approach to departure), but it appears to be very small. For simplicity we set $p_{min}=0$. Therefore, if not otherwise stated, we use $p_{min} = 0$ and $p_{max}=1$ as default parameters. We have checked that this simplification does not affect the qualitative results. Note that we cannot measure the global injury level $I$ directly, and the fraction of prey school covered by injuries is only a visual proxy for $I$. Therefore it is advantageous to express $p_c$ as a function of total number of attacks on the prey $n_a$. 

\begin{figure}[htb]
 \begin{center}
	 \includegraphics[width=0.49\textwidth]{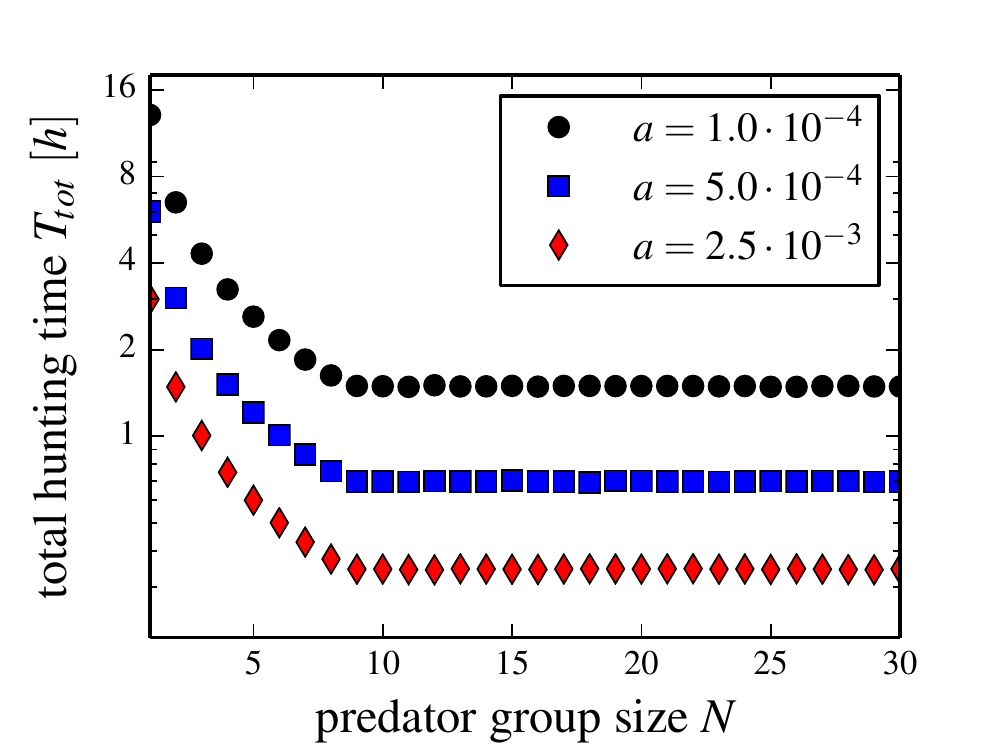}
 \end{center}
 \caption{Total hunting time $T_{tot}$ versus predator group size for different values of the $p_c$ growth rate $a$ (semi-logarithmic scale). The total hunting time decreases strongly for increasing predator group sizes ($N<1+\tau_a/\tau_r$), and then becomes independent of the group size $N$. It can vary from $~6-8h$ for a solitary hunter to $~1h$ for a groups larger than ten individuals. However, $T_{tot}$ depends strongly on the choice of $a$. The blue line (squares) shows $T_{tot}$ for parameters as used in the main text ($a=5\cdot10^{-4}$).  
Other parameters as in the main text: $p_{min}=0$, $\tau_a=2.6s$, $\tau_r=20s$, $S=200$.
\label{fig:SI_Ttot}
}
\end{figure}

\begin{figure}[htb]
 \begin{center}
	 \includegraphics[width=0.49\textwidth]{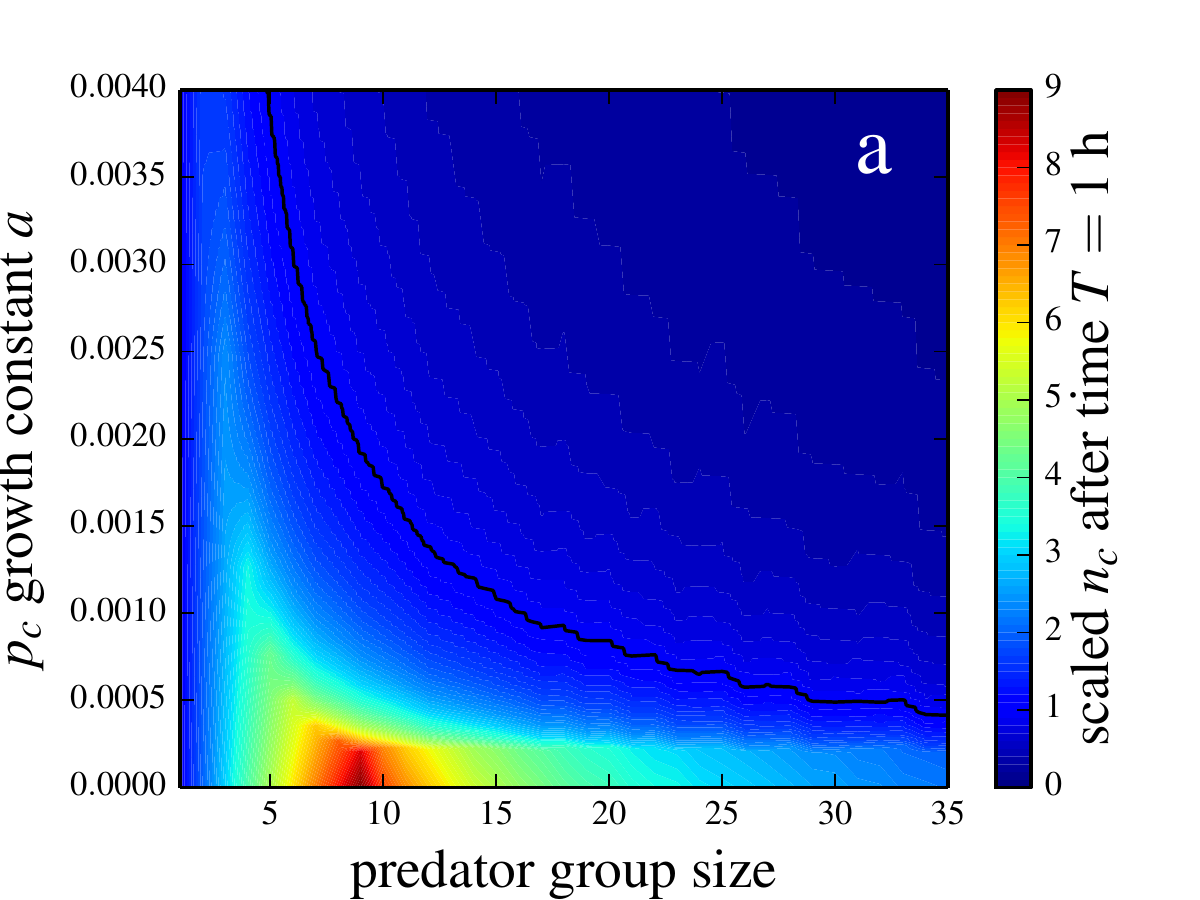}
	 \includegraphics[width=0.49\textwidth]{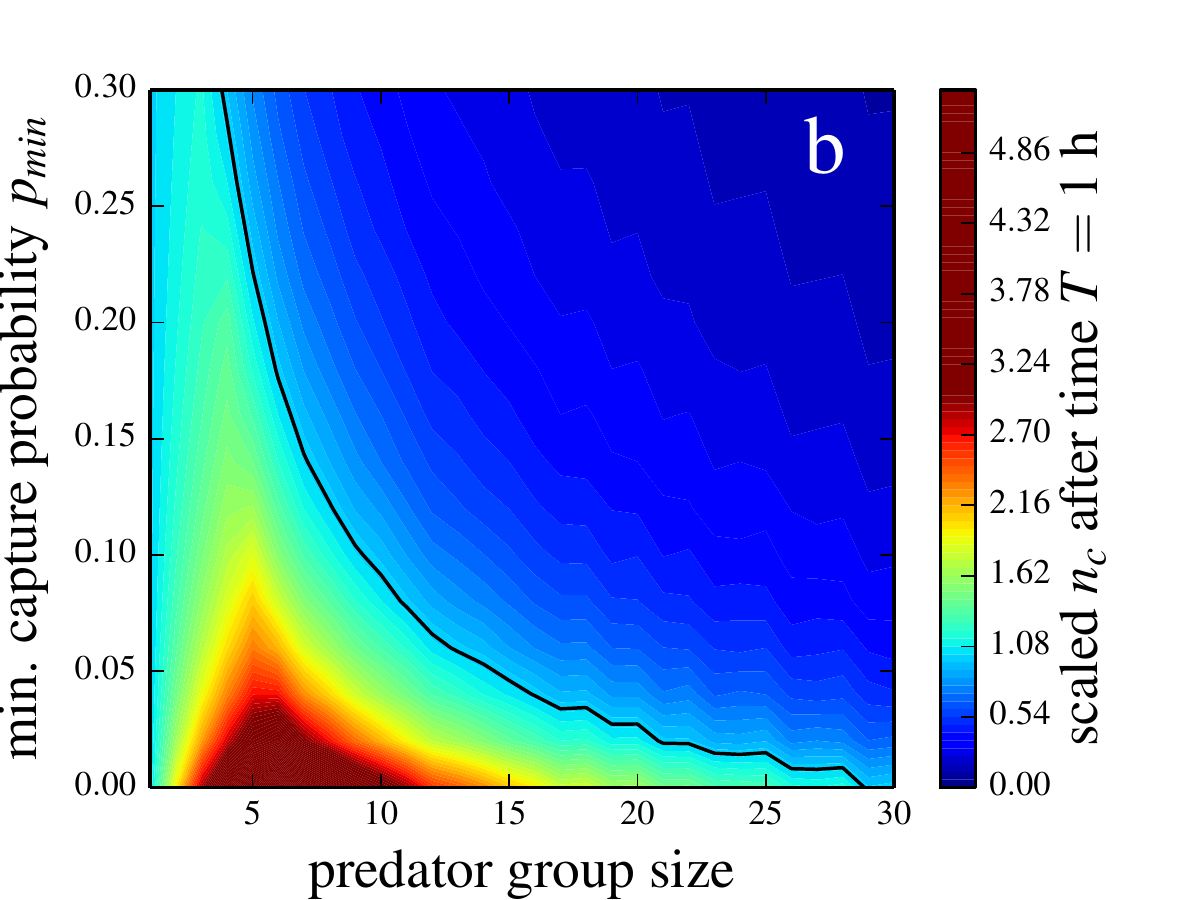}
 \end{center}
 \caption{(a) Number of prey captured after a time $T=1 h$ versus predator group size for varying the growth rate of the capture probability $a$  and (b) the min. capture probability $p_{min}$. $n_c$ is normalised by the number of prey captured by a solitary predator. The second $y$-axis shows average  capture probability per attack for a solitary hunter $\langle p_c \rangle_s$ calculated from the number of attacks of a solitary hunter required to catch 100 fish. Solid lines show the contours corresponding to the value of $n_c=1$ (same as solitary hunter) and therefore represent the border of the region where group hunting is beneficial. Default parameters as in the main text (if not varied): $p_{min}=0$, $a=5\cdot10^{-4}$, $\tau_a=2.6$, $\tau_r=20$. 
\label{fig:SI_nc_vs_a_pmin}
}
\end{figure}

The increment in the injury level $\Delta I$ per attack may in principle be an arbitrary function of the global injury itself: $\Delta I=g(I)$. This in general implies a nonlinear dependence of $I$ on the number of attacks $n_a$: $I(n_a)=f(n_a)$. However, if we assume that the increment in the level of injury per attack is constant $\Delta I=const.$, then $I$ is simply proportional to $n_a$ ($I\propto n_a$). In this case we can set, without loss of generality,  $\Delta I=1$, rescale $a_I$ to a new constant $a$, and replace $I$ by $n_a$. The probability can then be directly expressed as a function of the number of attacks $n_a$ as:
%%%%%%%%%%%%%%%%%%%%% EQUATION %%%%%%%%%%%%%%%%%%%%%%%%%
\begin{align}\label{eq:pc_vs_na}
	p_c(n_a)  & =p_{min}+(p_{max}-p_\text{min})(1-e^{-a n_a})\ .
\end{align}
%%%%%%%%%%%%%%%%%%%%%%%%%%%%%%%%%%%%%%%%%%%%%%%%%%%%%%%
We have explored how the rate of increase of the capture probability, $a$, and $p_{min}$
affects whether group hunting is beneficial for individuals in groups (Fig. \ref{fig:SI_nc_vs_a_pmin}). The main observation is a decrease of the maximal beneficial hunter group size with increasing $p_{min}$. The choice of $a$ will strongly affect the overall time of the hunt before all prey individuals are captured and the average capture efficiency during the hunt (no. prey captured / no. attacks). Large values of $a$ yield high average capture efficiencies already after a $1h$ of hunting ($p_c>0.3$) and as a consequence very short hunts ($<1h$) for reasonable prey school sizes. On the other hand extremely low values lead to very low initial capture efficiencies ($p_c\ll0.1$) and eventually result in very long hunting times $>4h$ (see Fig. \ref{fig:SI_Ttot}). In combination with other parameters used, a choice of $a=5\cdot10^{-4}$ yields total hunting times which appear consistent with our observations: $T\sim2h$ for predator groups $>5$.

\subsection{\normalsize Nonlinear dependence of injuries on number of attacks}
\label{sec:nonlinear}

\begin{figure}[p]
 		%\begin{center}
		%\begin{picture}(100,100)
	%\put(-120,0){
	 \includegraphics[width=0.32\textwidth]{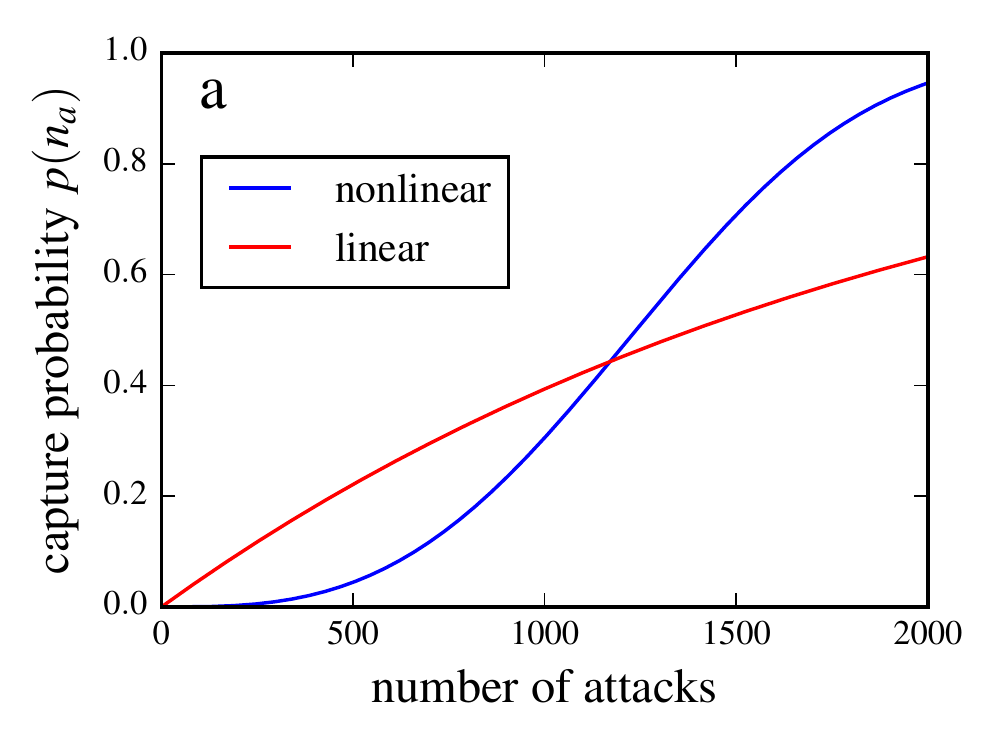}
	 \includegraphics[width=0.325\textwidth]{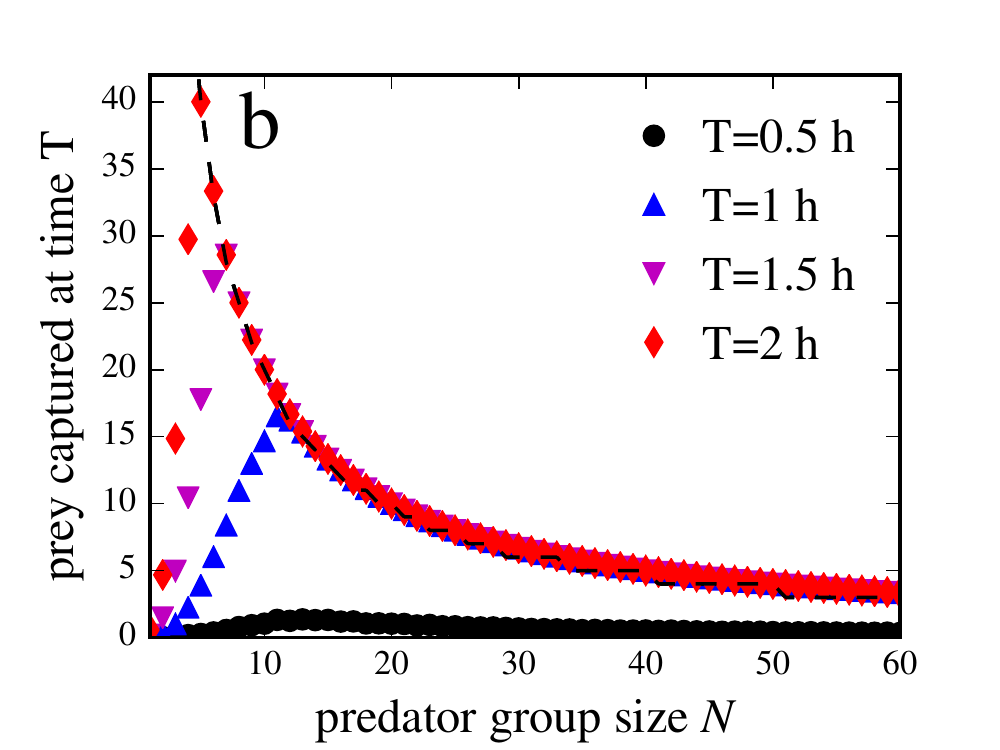}
	 \includegraphics[width=0.325\textwidth]{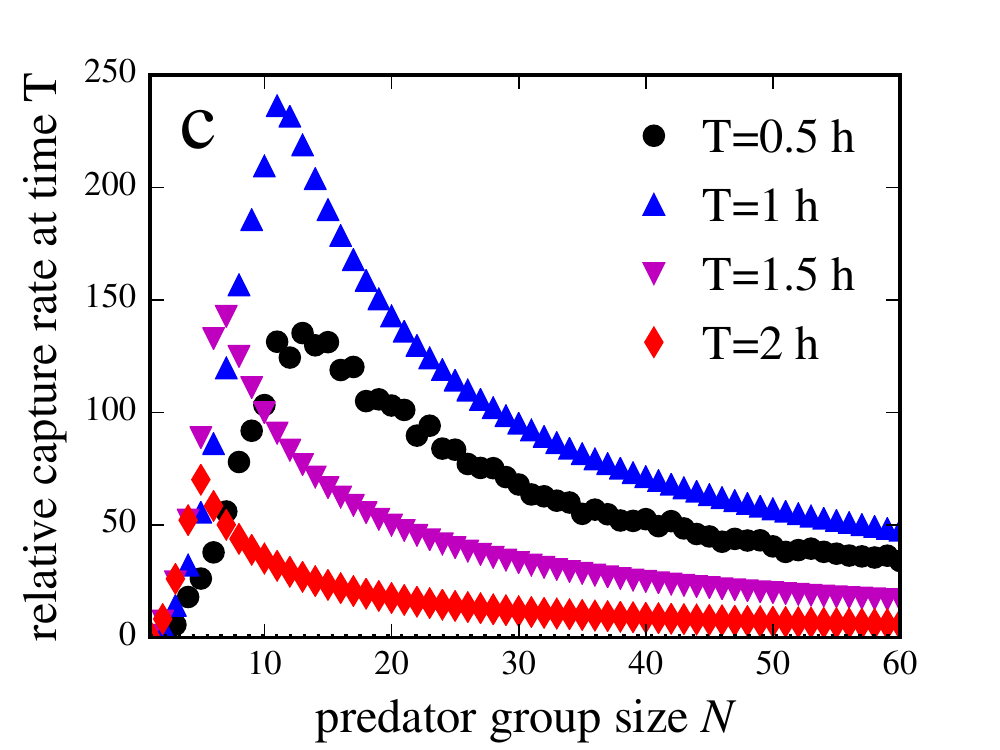}
 	%}
	%	\end{picture}
 		%\end{center}
    \caption{Nonlinear dependence of injury on number of attacks: (a) Comparison of the linear and nonlinear capture probability, where the factor $\beta$ of the nonlinear model were chosen to match the cumulative $p_{catch}$ for $n_a=2000$ ($\gamma=3$). (b) Absolute number of prey captured per hunter versus predator group size at different hunting times. (c) Relative capture rate versus groups size for different hunting times. All other simulation parameters as in the main text. \label{fig:capture_NL}
}
\end{figure}

In the main text, we assumed that the injury increment per attack is independent of the number of previous attacks, which implies a linear relationship between the number of attacks  $n_a$ and the prey injury level $I$. Here, we demonstrate that the qualitative results of our model remain unchanged for a nonlinear dependence of the injury level on $n_a$. We assume $I(n_a)=(\beta n_a)^\gamma$ with $\gamma>1$. A direct consequence of such a nonlinear dependence is a sigmoid shape of $p_{catch}(n_a)$, where the function changes from convex to concave (change in sign of the second derivative) at a finite number of attacks. The factor $\beta$ controls the location of the midpoint of the sigmoid ($p_{catch}=0.5$), wheras the exponent $\gamma$ determines the steepness of the sigmoid. In order to be able to compare results obtained for the linear and nonlinear model variants, we choose the additional nonlinear parameters so that the cumulative capture probability $\sum_{j=0}^{n_a} p_{catch}(j)$  for $n_a=2000$ attacks is approximately the same for both the nonlinear and  linear model variant (see Fig. \ref{fig:capture_NL}a). 

\begin{figure}[p]
 		%\begin{center}
		%\begin{picture}(100,100)
	%\put(-120,0){
	 \includegraphics[width=0.35\textwidth]{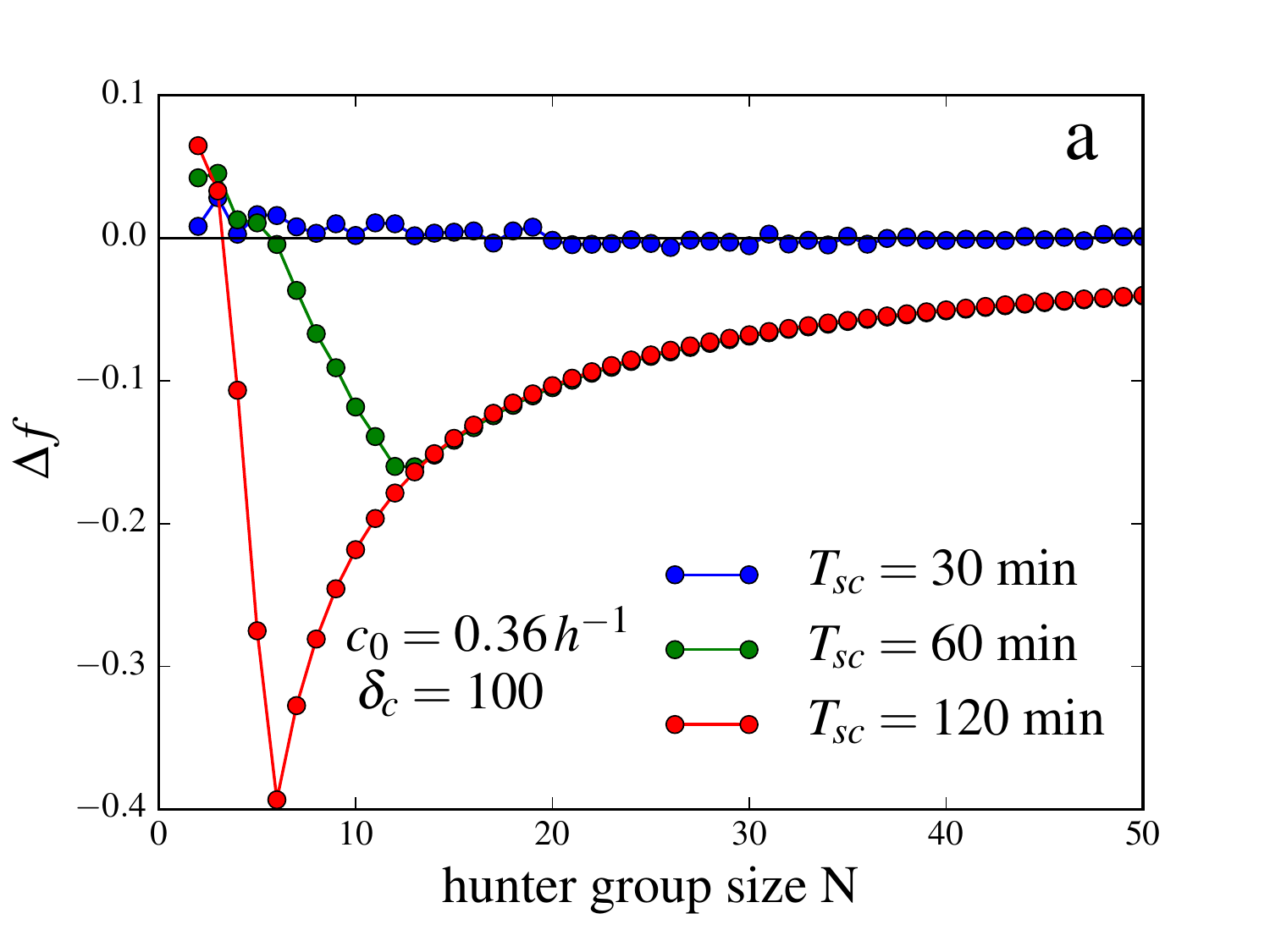}
	 \includegraphics[width=0.35\textwidth]{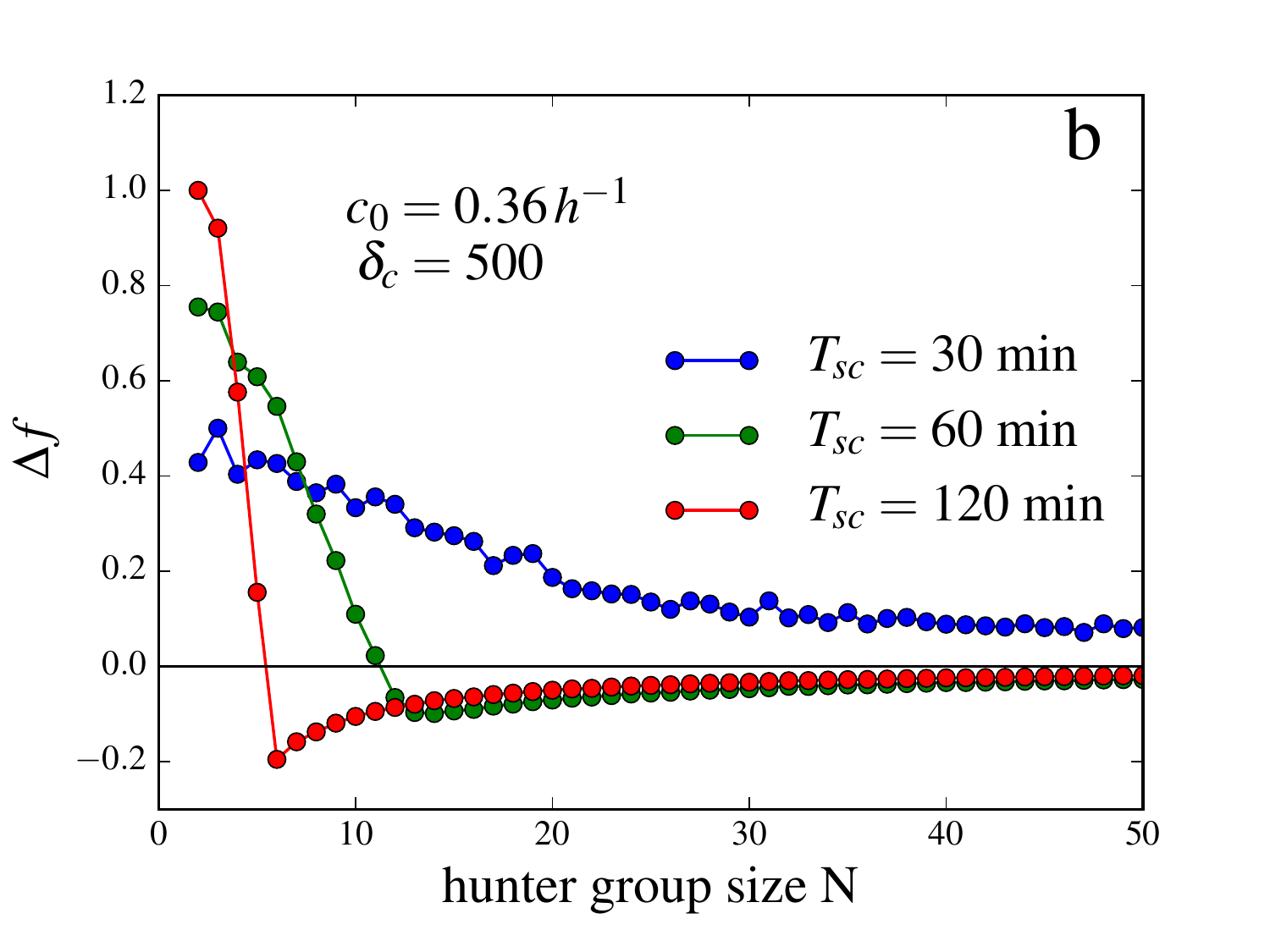}
	 \includegraphics[width=0.29\textwidth]{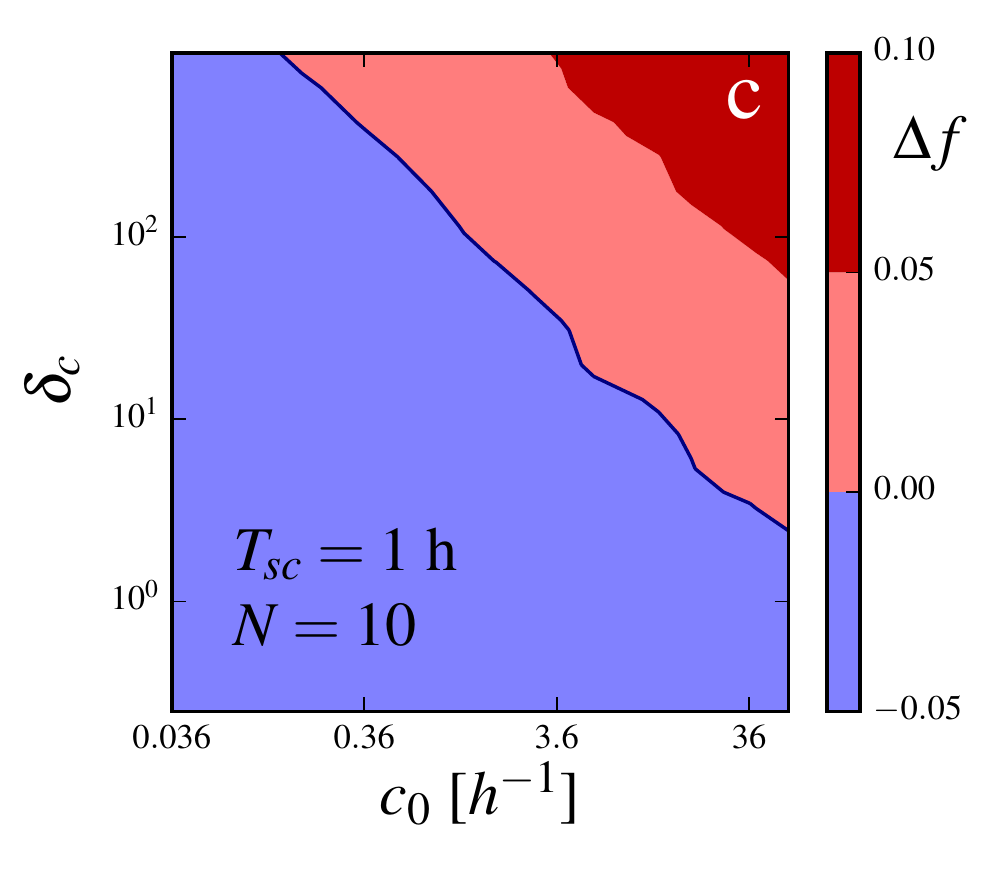}
 	%}
	%	\end{picture}
 		%\end{center}
    \caption{Nonlinear dependence of injury on number of attacks:: Relative fitness difference $\Delta f$ with $c_0=0.36 \text{h}^{-1}$ and different values of the relative energetic costs of attacks (a) $\delta_c=100$, and (b) $\delta_c=500$. (c) $\Delta f$ versus $c_0$ and $\delta_c$ for fixed $T_{fr}=0.5$h and $N=10$ (blue region indicates $\Delta f<0$; i.e. where free riding is not beneficial). Parameters: $p_{int}=0$ (no interrupt), $\gamma=3$ and $\beta$ chosen accordingly to match the cumulative $p_{catch}$ for $n_a=2000$ of the linear model; all other simulation parameters as in the main text. \label{fig:scrounging_NL}
}
\end{figure}

In general, the nonlinear injury dependence results in a lower capture probability at low number of attacks in comparison to the linear case. This situation reverses for large $n_a$ as $p_{catch}$ increases strongly in the vicinity of the sigmoid midpoint before asymptotically approaching $p_{catch}=1$ (Fig. \ref{fig:capture_NL}a). As a consequence we expect in general lower capture rates at short times for all group sizes, as well as larger potential benefits from free riding at short hunting times. This is confirmed by the corresponding simulation results: First, we observe a very low capture success for small hunting group sizes and short hunting times (low accumulated number of attacks) as shown in Fig. \ref{fig:capture_NL}b. In particular, this has a strong effect on the relative capture rate, normalised by the rate of solitary hunters. We observe a strong increase in the effective capture rates at short hunting times for group hunting, with respect to the number of prey a solitary hunter would have caught under the same conditions (Fig. \ref{fig:capture_NL}c). Second, in the nonlinear case (see main text and Supplementary section 2.4 \& 2.5), we observe an increased fitness benefit for free riders. This becomes particularly prominent at short hunting times (Figs. \ref{fig:scrounging_NL}a,b) and yields a larger region in the energetic parameter space where free riding appears beneficial (Fig. \ref{fig:scrounging_NL}c). However, our model predicts that also in the nonlinear case, free riding is unlikely to give fitness benefits for reasonable energetic parameters (in particular $c_0\ll0.5h^{-1}$).

\subsection{\normalsize Stochastic model with random attack and preparation times}

\begin{figure}[htb]
 \begin{center}
	 \includegraphics[width=0.49\textwidth]{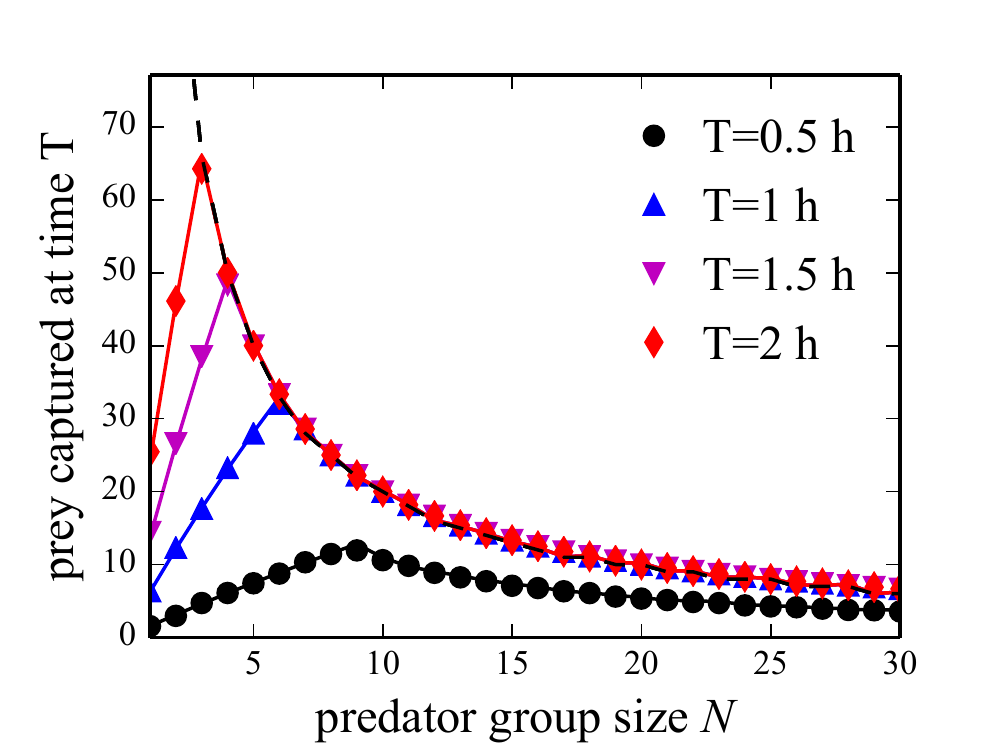}
 \end{center}
 \caption{The number of fish captured by a single hunter as a function of group size $N$ at
different times, $T$, available for hunting, for parameters consistent with our experimental observations (for details on parameter choice see above). For short hunting times, capture success increases up to a maximum and then decreases monotonically with increasing $N$. For longer times and large $N$, the curves collapses on
the limiting line set by the average number of prey per predator. Solid lines represent prediction of Eq. 1 taking into account the upper limit given $S_0/N$ shown by the dashed line. Symbols represent the results of model simulation. Each point represents an average over 100 independent runs. 
\label{fig:SI_nc_unscaled}
}
\end{figure}

\begin{figure}[htb]
 \begin{center}
	 \includegraphics[width=0.49\textwidth]{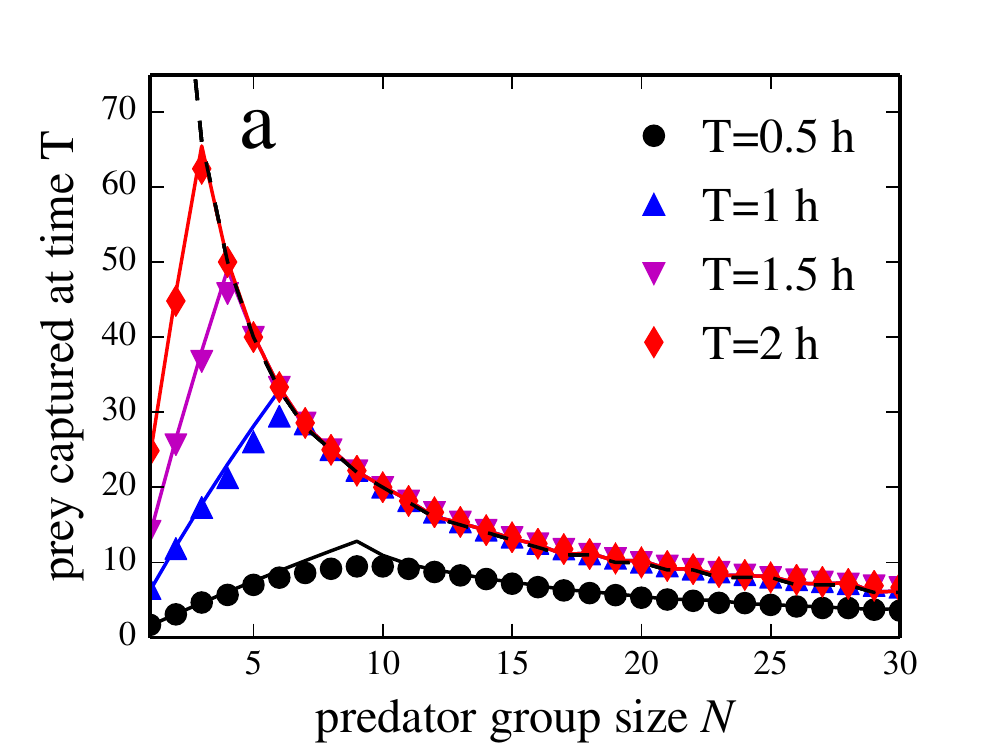}
	 \includegraphics[width=0.49\textwidth]{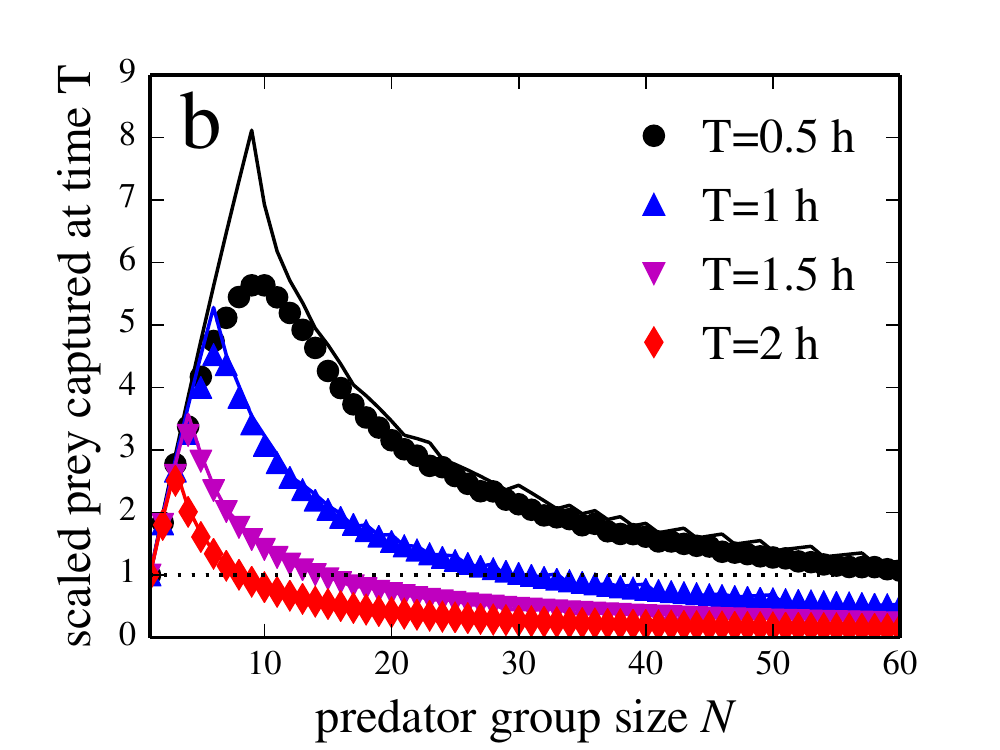}
 \end{center}
 \caption{
Group hunt model with stochastic attack and preparation times: (a)  Number of prey captured per
individual versus predator group size at different times available for hunting.
Solid lines represent the prediction of Eq. 1 (main text) taking into account the upper limit given
$S_0/N$ shown by the dashed line. Symbols represent the results of model simulation. 
Each point represents an average over 100 independent runs. 
(b) Number of prey captured per individual as a function of N scaled by the number of prey captured for a solitary predator (horizontal dotted line). The largest group sizes, which offer an advantage
to solitary hunting are typically observed for short times $T \leq 1h$ and
decreases for large times (or small prey schools). For details on the modified model see section 1.3 above. Model parameters as in the main text: $\tau_a = 2.6s$, $\tau_r =20s$,  $a =5\cdot10^{-4}$, $p_{min} =0$, $p_{max}=1$, $S_0 = 200$.
\label{fig:SI_stoch_mod}
}
\end{figure}

In the main text we assume fixed, constant attack and preparation times. This yields perfect turn-taking of individuals with the order of attacking individuals set by the initial condition, which are randomised for each hunt. However, a perfect turn-taking behaviour may be questioned from a biological point of view. Furthermore, in general, deterministic temporal sequences may lead to pathological behaviour of mathematical models for certain parameter combinations (``resonances''). In order to confirm that our results are robust and independent on the turn-taking behaviour, we extended the simple model discussed in the main text, to random attack and preparation durations: Instead of fixed duration, we model the attack and preparation times as random variables $t_a$ and $t_r$ drawn from an exponential distribution with averages $\tau_a$ and $\tau_r$, respectively. The initial order of attacking hunters is again random as in the main text, but now the order of the hunters within a single run does not persist but changes randomly due to the stochasticity of the attack and preparation durations. 
Figure \ref{fig:SI_stoch_mod} shows the comparison between the simulation results for the (scaled) number of prey captured for this modified model with our theoretical predictions (compare to main text, Fig. \ref{fig:model}b and Fig. \ref{fig:SI_nc_unscaled}). All simulation results were obtained by averaging over 100 independent simulations. In particular for longer hunts ($T$ large), only the average waiting times are relevant and the results of the fully stochastic model strongly match our theoretical predictions. For short times $T$, the additional stochasticity of the hunting process leads to smoothening of the maximum number of prey captured, but the position of the maximum and the maximal group size beneficial for hunting remain essentially unchanged. 

\subsection{\normalsize Energy Balance Equation}
As the detailed metabolic costs of swimming and attacking in sailfish are not documented, we considered a simple, yet generic model, where different energetic costs and benefits are summarised in a few key parameters. We assume that each hunter has a base rate of energy expenditure $C_0$, which includes all metabolic costs of swimming required to stay with the prey school, but excludes any additional energetic investment required to perform an actual attack sequence. For simplicity, we assume that attacks and captures are instantaneous events, which happen at discrete points in time $t_a$ and $t_c$. This is a reasonable approximation as in general the average attack time $\tau_a$ will be much shorter than the total hunt time $T_h$. Prey capture is only possible during an attack, thus capture points $t_c$ are always a subset of the attack points $t_a$. The additional costs of attack are included into the model as a constant energy decrement due to an increased energy consumption rate during an attack $\Delta E_{a}=C_a\tau_a$. The energy benefit from each captured prey is given by a constant increment $\Delta E_{c}$.  Thus the total energy $E_i$ for an individual hunter $i$ during a hunt evolves according to the following balance equation:
\begin{align}
	\frac{dE_i(t)}{dt} 	&=-C_0-\Delta E_{a}\delta(t-t_{a})+\Delta E_{c}\delta(t-t_{c}) \\
				&=-C_0-C_a\tau_a\delta(t-t_{a})+\Delta E_{c}\delta(t-t_{c}) 
\end{align}
with $\delta(t-t')$ being the Dirac delta function. Without loss of generality, we rescale all terms by the energy increment due to prey capture $\Delta E_c$, thus we measure the energy in units of the average energy content of a single prey item. The rescaled equation reads:
\begin{align}
	\frac{de_i(t)}{dt}=-c_0-c_{a}\tau_a\delta(t-t_{a})+\delta(t-t_{c})
\end{align}
with $e_i=E_i/\Delta E_{c}$, $c_x=C_x/\Delta E_c$. 

By integrating over the entire time of the hunt $T_{h}$, we obtain the overall energy payoff per individual as:
\begin{align}
	\Delta e_{total,i} = -c_0 T_h - c_a \tau_a n_{a,i} +  n_{c,i}
\end{align}
with $n_{a,i}=\int_{t_0}^{t_0+T_h} \delta(t-t_a)dt $ and $n_{c,i}=\int_{t_0}^{t_0+T_h} \delta(t-t_c)dt$ being the number of attacks and the number of prey captured by the focal individual. 

Finally we rewrite the energy payoffs by pulling out the base rate $c_0$ to obtain:
\begin{align}
	\Delta e_{total,i} = -c_0 (T_h + \delta_c \tau_a n_{a,i}) + n_{c,i}
\end{align}
Here the dimensionless number $\delta_c=c_a/c_0$ represents the effective increase of the energy consumption during an attack relative to the base consumption rate. A value of $\delta_c=1$, would correspond to doubling of the energy consumption rate during an attack sequence. We discuss only biologically relevant parameter values $c_0$ and $\delta_c$, where the average energy pay-offs are positive $\langle \Delta e_{total} \rangle>0$. All other model parameters are the same as in the main text. 

Because the routine metabolic rate of adult sailfish in unknown, we estimated a conservative value of $c_0$ based on the routine metabolic rate (RMR) of Blue-fin tuna (\textit{Thunnus orientalis}). The RMR of a 8.1 kg tuna is $\sim$ 280 mg O$_2$ $kg^{-1}$ $h^{-1}$ at 25 $^o$C \cite{blank2007temperature}. The length:weight relationship of adult sailfish is given by: Log W = $-5.443 L^{3.007}$ where W is the mass and L is the length of a sailfish \cite{velayudham2012length}.  A sailfish 240 cm in length, therefore, has a mass of 51.8 kg.  We can scale the RMR of tuna according to the mass of the sailfish by the following scaling factor:  $RMR_{sailfish} = RMR_{tuna} (8.1/51.8)^{(1-0.8)}$ \cite{schurmann1997effects,edwards1972experimental}. This gives a RMR for a sailfish as 193.2 mg O$_2$ $kg^{-1}$ $h^{-1}$.  This equates to a sailfish requiring 240 g O$_2$ per day.  Given the oxycalorific coefficient is 13.59 \cite{brett1979physiological}, this equates to a sailfish requiring 3263 kJ of energy per day to maintain RMR.  Domenici et al. (2014) found that the average length of sardines found in a sailfish's stomach was 19 cm  \cite{domenici_how_2014}. This gives a mass of 57 g per sardine, based on their length:weight relationship \cite{fishbase}. The energy content of similar species (\textit{Sardinops melanostictus} and \textit{Clupea harengus pallasi}) is $\sim$ 6 - 9.6 kJ g$^{-1}$ \cite{benoit2004prey,anthony2000lipid}. An individual sardine, therefore, may provide 342 - 547 kJ of energy to a sailfish. Combining this information together, we estimate that a sailfish would require between 6 - 9.5 sardines per day to maintain RMR.  This equates to $\sim$ 0.25 - 0.4 sardines per hour. %Given that our assumptions are based on the RMR of a fish swimming slowly, increased activity by the sailfish may increases these values by a factor of $\sim$ 2 - 5. 
Here we choose a value of $c_0$ = 0.0001, meaning that a sailfish would need to eat 1 sardine every 10000 seconds (2.8 hours, or 0.35 sardines per hour) to balance energy intake and expenditure. We note that with decreasing values of $c_0$, the free riding strategy becomes increasingly unlikely. 

Another key question is; what is the realistic range of values for $\delta_c$? Our earlier observations indicate that the speed of approach towards the prey school during an attack sequence is similar to the continuous swimming of non-attacking individuals \cite{marras2015not,domenici_how_2014}. However, sailfish sometimes initiate rapid swimming bursts when the school attempts to escape into the depths or when chasing single prey that have left the school. Thus significant additional energetic costs of an attack sequence can only originate from these burst swims, the slashing motion of the bill, turning, and prey handling, which are likely to only take-up a small fraction of the entire attack sequence. Even in the case of extremely high costs of slashing/capture, the relative increase in costs of an attack are most likely of the order of the base rate energy consumption ($\delta_c \sim 1-10$). It seems unlikely, therefore, that a free riding strategy could yield benefits to individuals.

\subsection{\normalsize Free riding during interrupted hunts}
We also take into account the possibility of the hunt being interrupted due to external influences, for example, the arrival of other predators (e.g. dolphins). This is modelled by a constant probability $p_{int}$ of the hunt being terminated. For $p_{int}=0$, no interruption takes place and the hunt continues until all prey are captured. For finite $p_{int}>0$ the hunt is interrupted randomly, and times available for hunting are exponentially distributed with the average time $T_h=1/p_{int}$. 

\begin{figure}[p]
 		%\begin{center}
		%\begin{picture}(100,100)
	%\put(-120,0){
	 \includegraphics[width=0.325\textwidth]{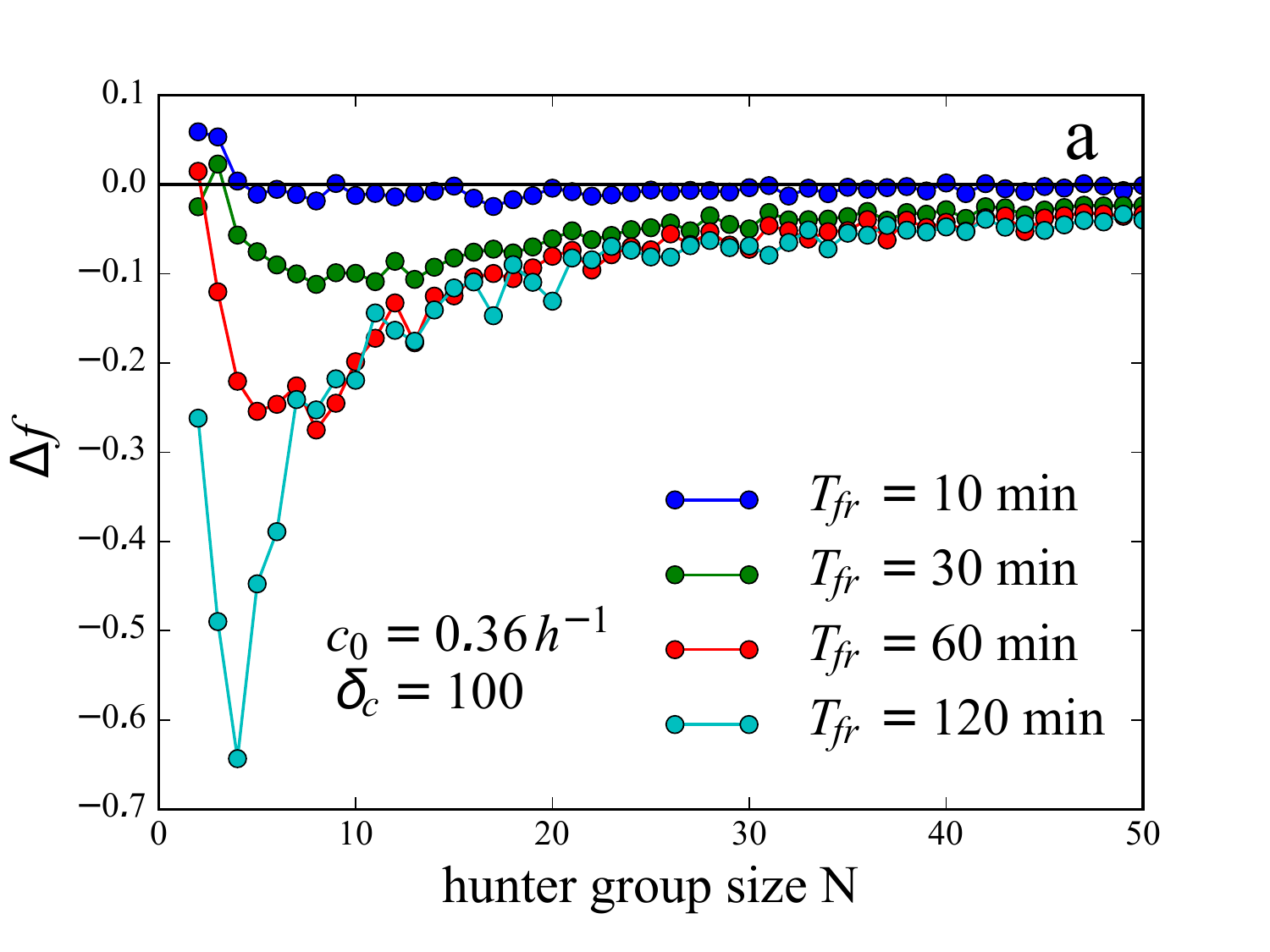}
	 \includegraphics[width=0.325\textwidth]{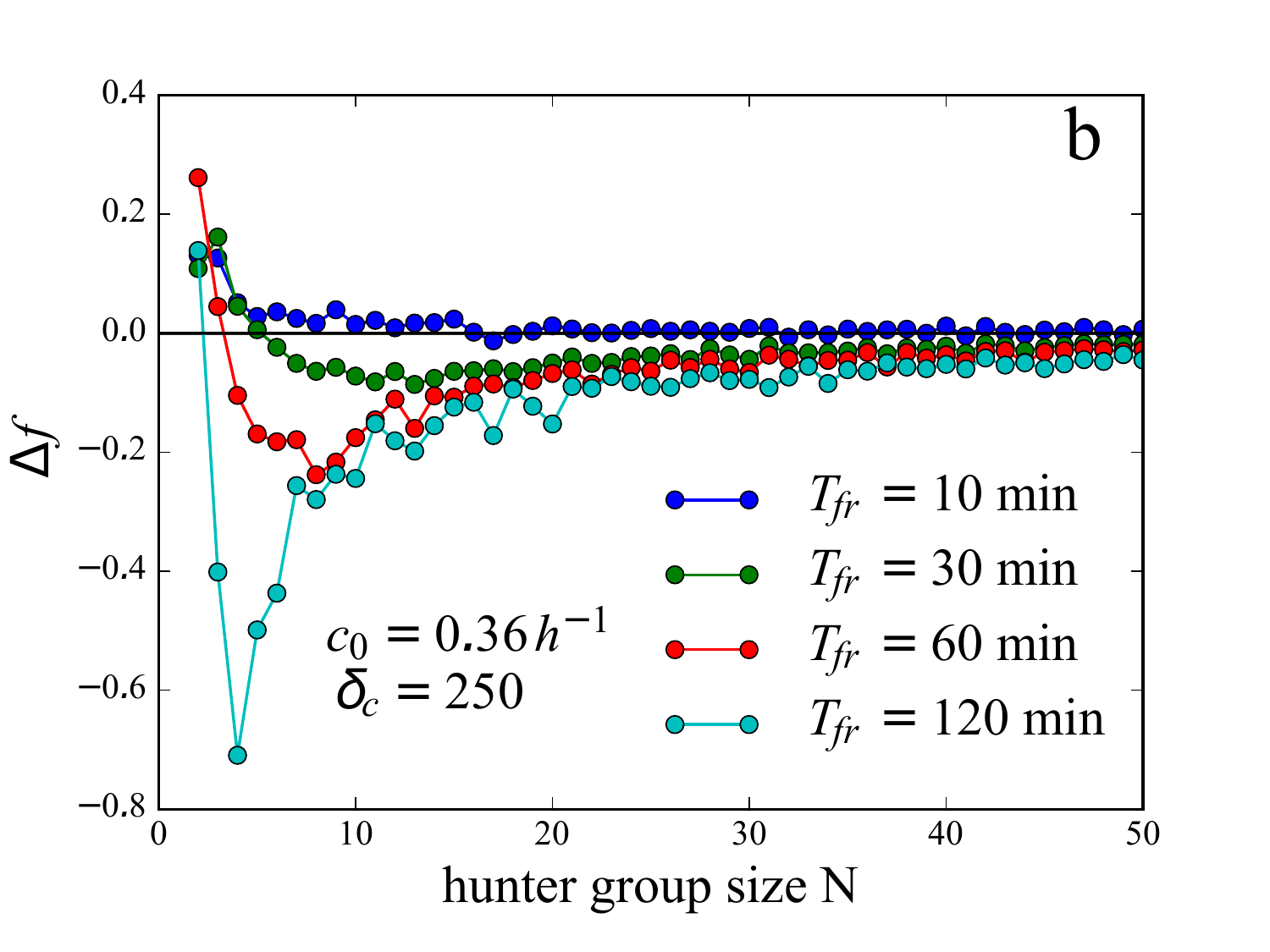}
	 \includegraphics[width=0.325\textwidth]{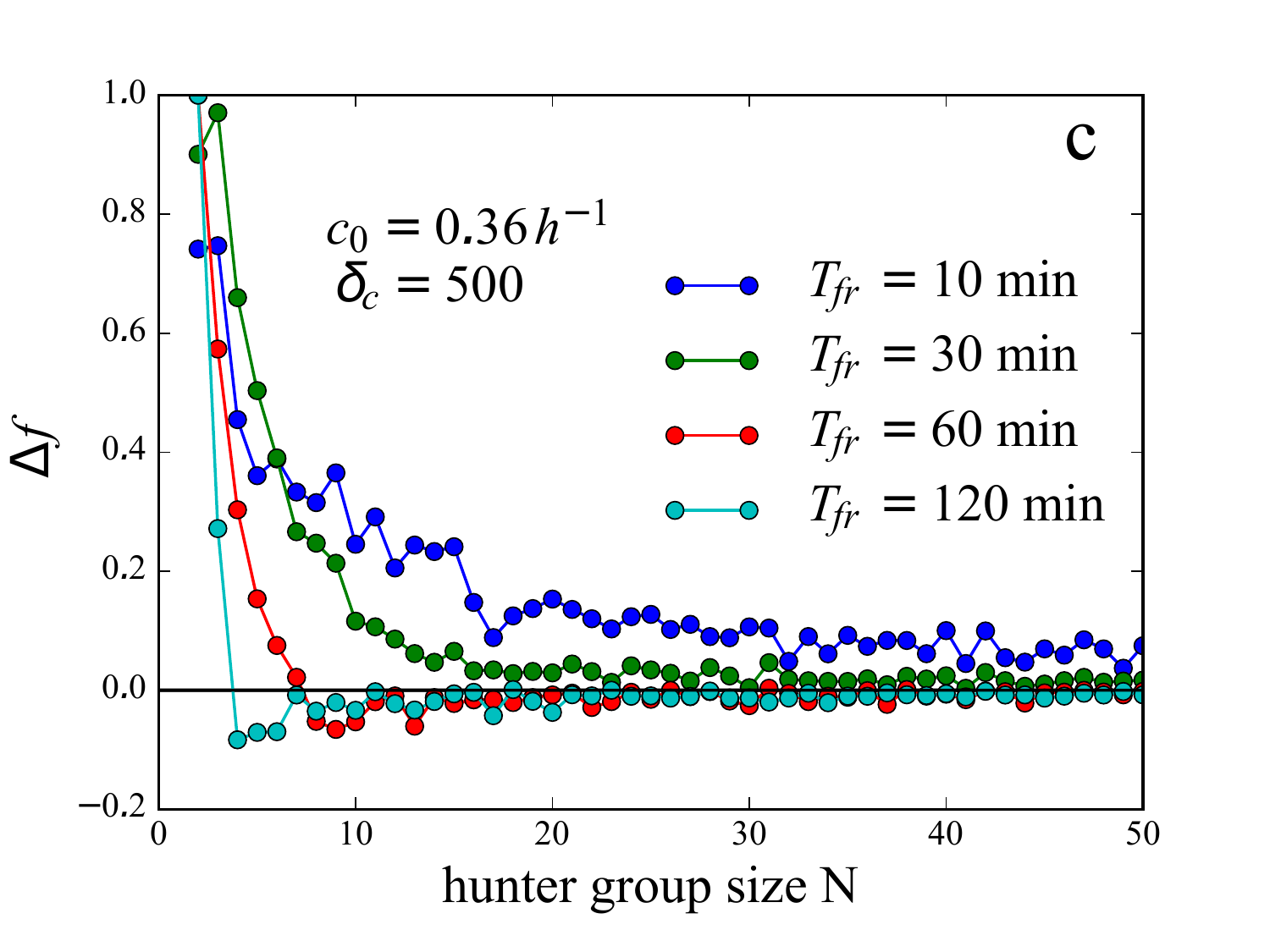}
 	%}
	%	\end{picture}
 		%\end{center}
 \caption{Relative fitness difference $\Delta f$ for the extended model with the possibility of interruption of the hunt with $p_{int}=0.0001$, with $c_0=0.36 \text{h}^{-1}$ and different values of the relative energetic costs of attacks $\delta_c=100$ (a), $\delta_c=250$ (b) and $\delta_c=500$ (c). All other simulation parameters as in the main text. \label{fig:scrounging_int}
}
\end{figure}

The qualitative findings do not depend on this model extension, and a significant energetic advantage of free riding can be observed only for very large $\delta_c$ values (Fig. \ref{fig:scrounging_int}). Interestingly, in these extreme cases with the finite probability of interruption of the hunt, the free rider advantage becomes more pronounced. This can be understood from the high energetic costs and negligible payoffs for producers in cases where the hunt was interrupted at an early stage.

%\end{description}
%\appendix

%\bibliographystyle{procb}
%\bibliography{SailfishHunting}

%\bibliographystyle{../Myapsrev}	
%\bibliography{SailfishManuscript}
\clearpage

\section{Supplementary Figures}

%Figure S1
\begin{figure}[htb]
 \begin{center}
	 \includegraphics[width=\textwidth]{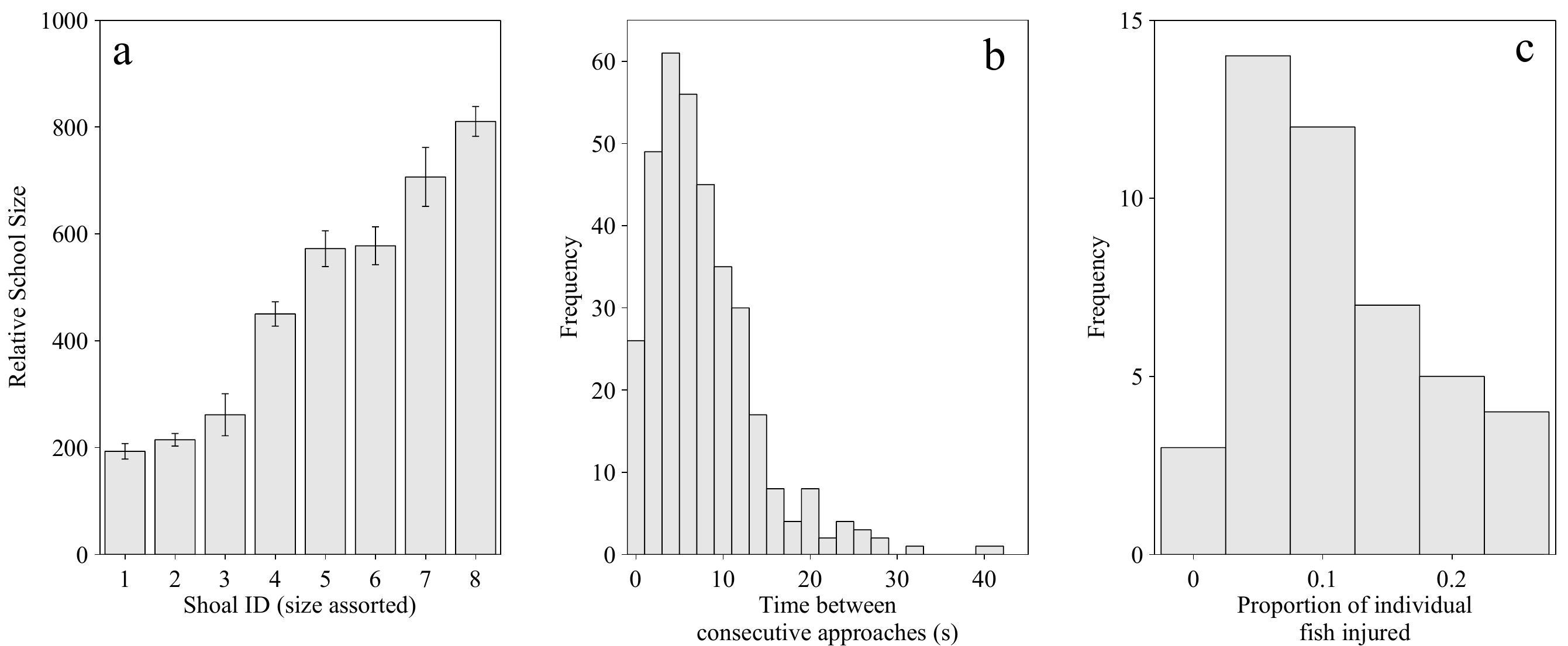}
 \end{center}
 \caption{Empirical Data (a) Proxy of the sizes of the sardine schools we analysed. School 1 was composed of approximately 25 fish, whereas
school 8 consisted of approximately 100-150 fish. Schools are ordered from smallest to largest (b) Distribution of the times between consecutive approaches by
different sailfish towards the sardine schools. (c) The maximum proportion of
injury on the bodies of individual sardines. 
\label{fig:SI_expdata}
}
\end{figure}

%Figure S3
\begin{figure}[htb]
 \begin{center}
	 \includegraphics[width=0.49\textwidth]{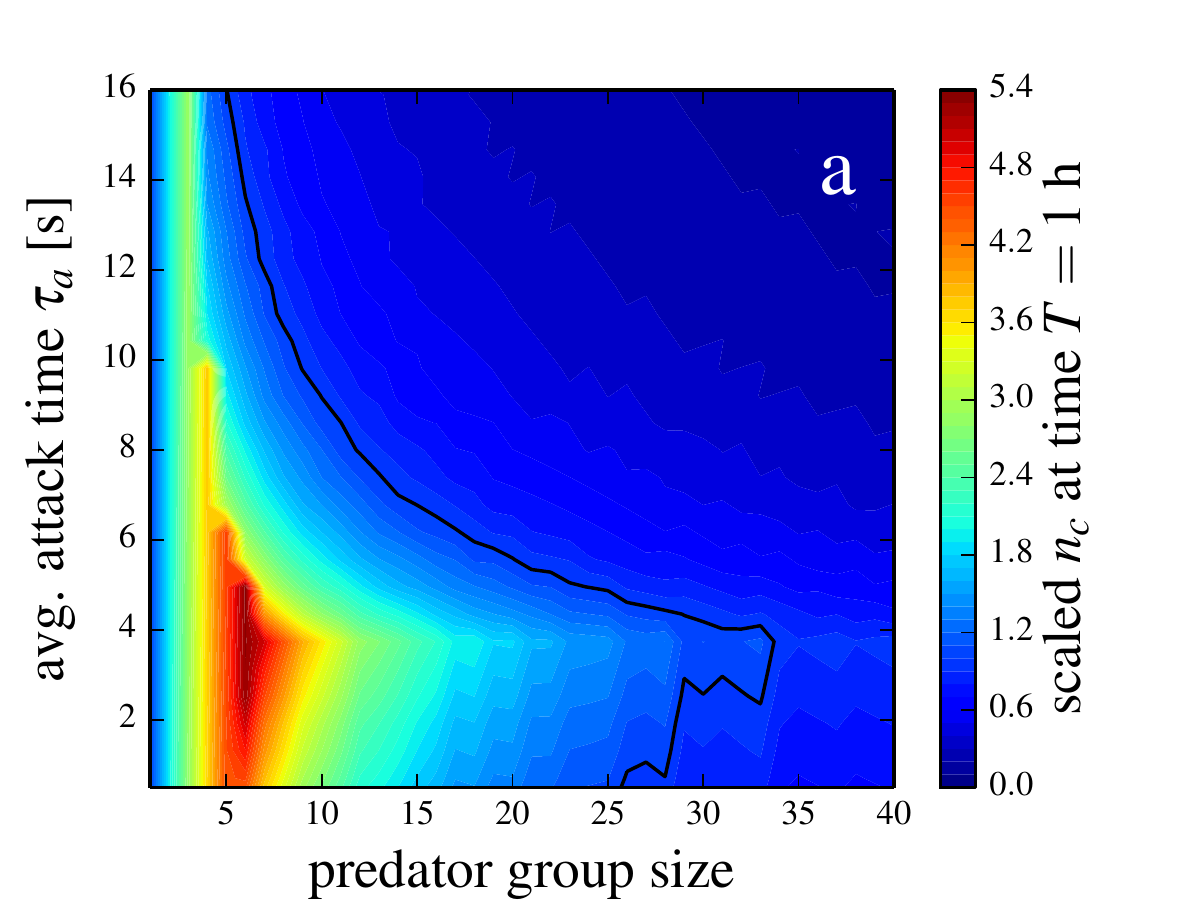}
	 \includegraphics[width=0.49\textwidth]{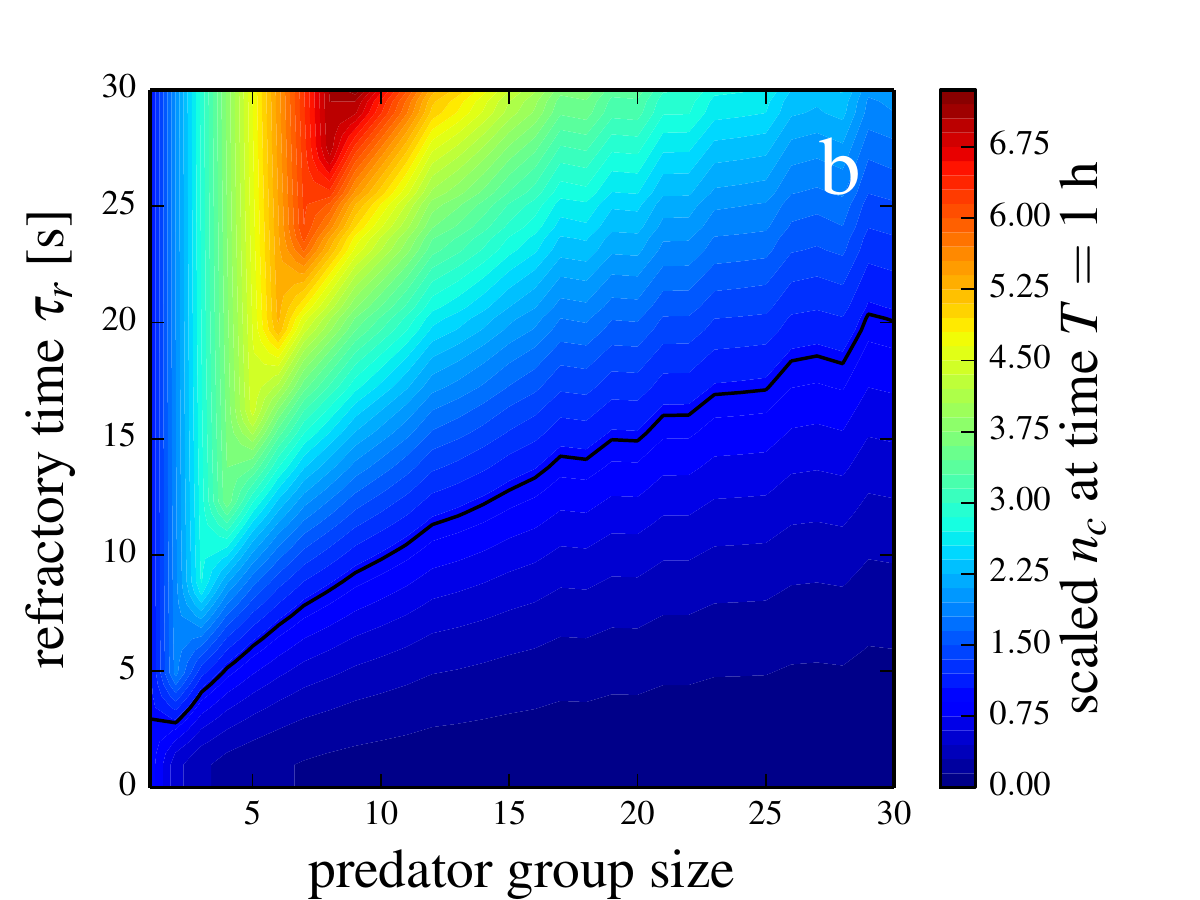}
 \end{center}
 \caption{(a) Number of prey captured after time $T=1 h$ versus predator group size for varying the attack time $\tau_a$  and (b) varying the refractory time $\tau_r$. $n_c$ is normalised by the number of prey captured by a solitary predator.  Solid lines show the contours corresponding to the value of $n_c=1$ (same as solitary hunter) and represent therefore the border of the region where group hunting is beneficial. 
Default parameters as in the main text (if not varied): $p_{min}=0$, $a=5\cdot10^{-4}$, $\tau_a=2.6$, $\tau_r=20$, $S=200$.
\label{fig:SI_nc_vs_ta_tr}
}
\end{figure}

%Figure S2
\begin{figure}[htb]
 \begin{center}
	 \includegraphics[width=0.49\textwidth]{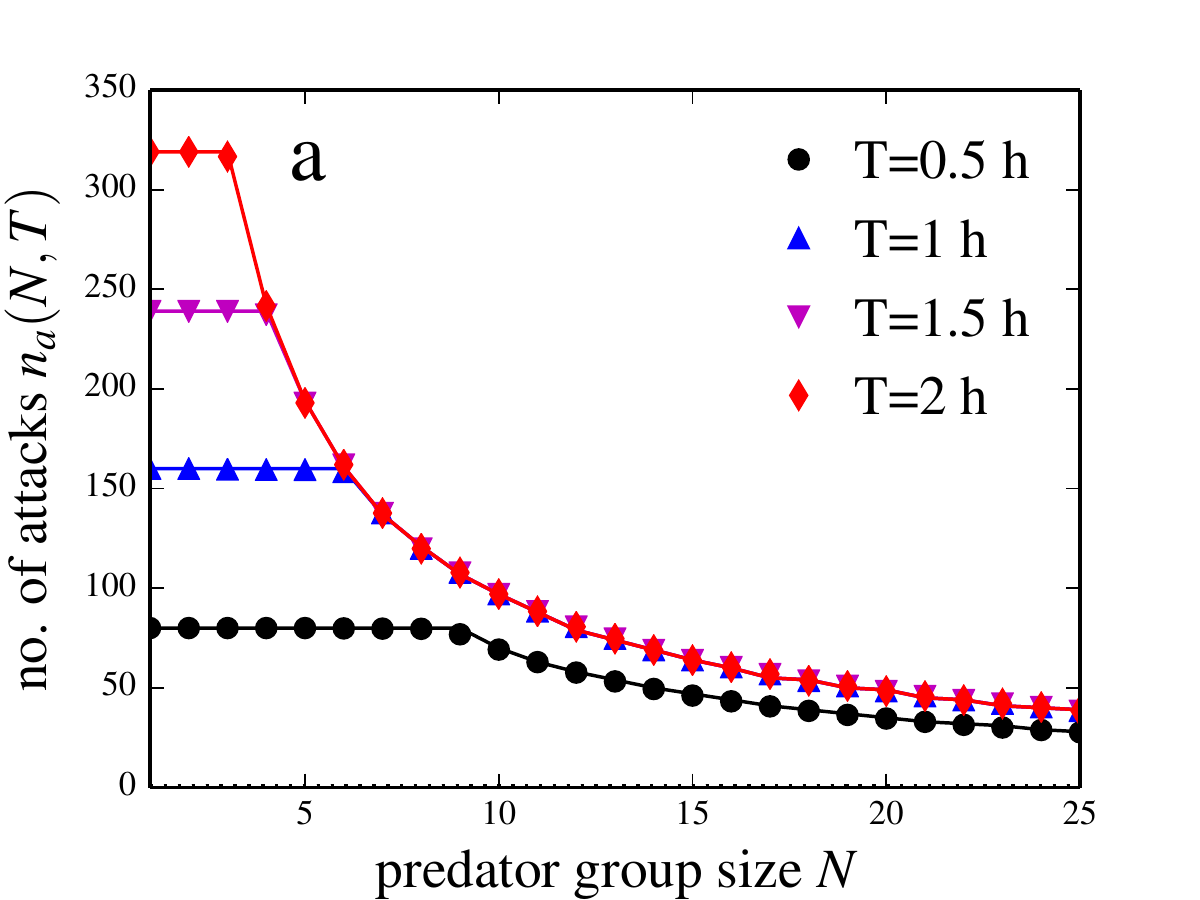}
	 \includegraphics[width=0.49\textwidth]{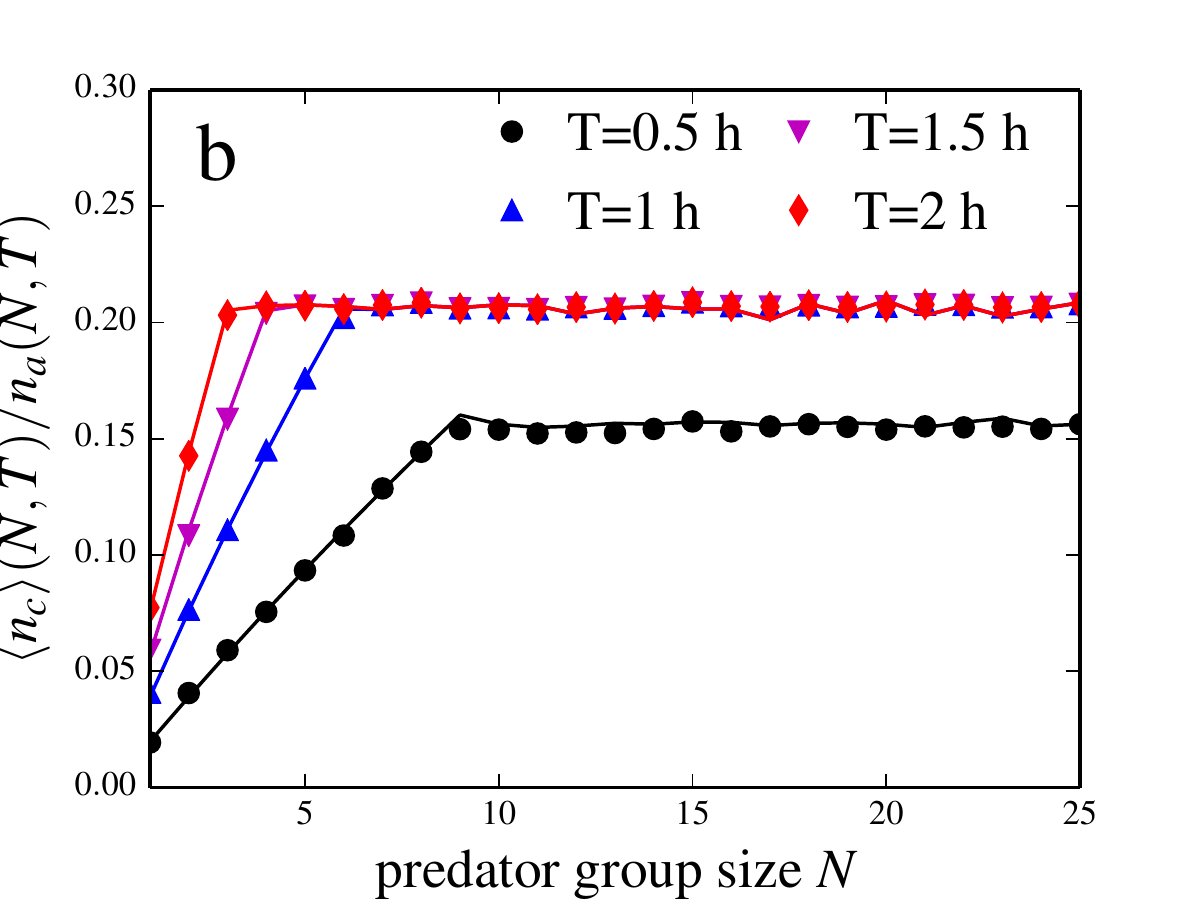}
 \end{center}
 \caption{(a) Number of attacks per predator versus group size and (b) average number of prey captured per attack versus group size at different times $T$.  Solid lines represent analytical predictions, whereas symbols represent results of model simulations. The number of attacks remains constant for $N<1+\tau_r/\tau_a$ - no temporal penalty of group hunting. For larger groups the number of attacks per individual decreases with group size as individuals have an increased idle time, where they have to wait until others perform their attacks. For all times, the average number of prey captured per attack increases initially with group size until $N=1+\tau_r/\tau_a$ and reaches a plateau for larger group sizes, as $\langle n_c \rangle$ (Fig. S5) scales in the same way as $n_a(N,T)$ (left) for large $N$. 
 Parameters as in the main text: $p_{min}=0$, $a=5\cdot10^{-4}$, $\tau_a=2.6s$, $\tau_r=20s$, $S=200$.
\label{fig:SI_model_na_nc}
}
\end{figure}

%Figure S4

%Figure S5

%Figure S6

% Figure

% Figure

%Figure S7

%Figure S8 

\clearpage
\section{Supplementary Movies}
Movie S1: One sailfish approaches and attacks a school of approximately 25 sardines.
When the sailfish's bill makes contact with the sardines, the sardines'
scales are removed. Note the injury marks on the sardines' bodies. The video is played at 1/8 of real time.  
\newline
\newline
Movie S2: Sequence demonstrating that sailfish alternate attacks on their prey. Notice not all approaches results in attacks.  Multiple prey are injured in the attack at 12 seconds, while only one sardine is caught.  Not all attacks result in prey capture, for example, see attack at 27 seconds.  Not all attacks result in prey being injured (attempted slash at 33 seconds).  Also notice that sailfish tend to abandon attacks if they get out of position, or if another sailfish approaches the school at the same time.   Video is played in real time.
\newline
\newline
Movie S3:  An injured sardine break aways from the school and is quickly captured by a sailfish. The video is played at 1/8 of real time.

\section{Supplementary Tables}

\begin{table}[!htb]	
	\begin{tabular}{|c|r|r|}
	\hline 
	School & \#Images: School Size & \#Images: Proportion of injuries \\
	\hline
	1 & 12 & 1 \\
	2 & 18 & 14 \\
	3 & 11 & 4 \\
	4 & 10 & 1 \\
	5 & 9  & 1 \\
	6 & 13 & 1 \\
	7 & 10 & 8 \\
	8 & 40 & 9 \\
	\hline
\end{tabular}
\caption{
Shoal identity and the number of images used to calculate school size or
the number of images uses to calculate the proportion of injuries in the school.
}
\end{table}

\begin{table}[!h]
\begin{tabular}{|c|l|}
	\hline
	Symbol & Description   \\ 
	\hline
	\multicolumn{2}{|l|}{Base Model} \\
	\hline
	$N$ & number of predators  \\
	$S_0$ & initial number of prey  \\
	$\tau_a$ &  average time required to perform an attack sequence  \\
	$\tau_r$ &  average time required to prepare for an attack sequence \\
	%$\tau_{r,min}$ &  minimal time required to prepare for an attack sequence \\
	$p_\text{min}$ &  capture probability at zero injury   \\
	$p_\text{max}$ &  maximal possible capture probability   \\
	$a$ & growth rate of injury/capture probability with each attack \\
	\hline
	\multicolumn{2}{|l|}{Model Extensions} \\
	\hline
	$c_0$ & metabolic base rate measured in number or prey items per unit time \\
	$\delta$ & increase of the energy consumption during an attack relative to the base rate \\
	$p_{int}$ & constant probability per unit time of the hunt being interrupted \\
	$\gamma$ & nonlinearity exponent for the dependence of injury levels on the number of attacks \\
	$\beta$ & nonlinearity factor \\
	\hline
\end{tabular}
\caption{
	Model parameters with description including parameters for the base model and the different extensions of the model discussed in the main text and supplementary information.
}
\end{table}

\clearpage
\bibliographystyle{procb}
\bibliography{SailfishHunting}

\end{document}